\documentclass[12pt]{article}
\usepackage{amsmath}
\usepackage{amssymb}
\usepackage{amsfonts}
\usepackage{amsthm,systeme}
\usepackage[usenames]{color}
\usepackage[skip=1ex]{caption}
\definecolor{AV}{rgb}{0.65,0.0,0}
\definecolor{GC}{rgb}{0,0.0,0.65}
\definecolor{WS}{rgb}{0,0.65,0}
\usepackage{graphics,epstopdf}
\usepackage[dvips]{graphicx}
\usepackage[margin=1.5cm]{geometry}
\usepackage{graphicx,subcaption,lipsum}
\usepackage[skip=-80pt]{subcaption}
\usepackage{subcaption}
\usepackage{tikz} 
\usepackage{multirow}
\usepackage{tabularx}
\usepackage{changes}
\newcolumntype{Y}{>{\centering\arraybackslash}X}

\usepackage[
      colorlinks=true,
      linkcolor=blue,
      urlcolor=blue,
      filecolor=blue,
      citecolor=blue,
      pdfstartview=FitV,
      pdftitle={},
      pdfauthor={},
      pdfsubject={},
      pdfkeywords={},
      pdfpagemode=None,
      bookmarksopen=true
]{hyperref}
\usepackage{float}
\usepackage{breqn}
\usepackage{subcaption}
\usepackage[section]{placeins}
\usepackage[nodisplayskipstretch]{setspace}
\newcommand{\beqs}{\begin{eqnarray}}
\newcommand{\eeqs}{\end{eqnarray}}

\usepackage{caption}
\begin{document}

\thispagestyle{empty}

\hfill{}

\hfill{}

\hfill{}

\vspace{32pt}

\begin{center}

\textbf{\Large Particle dynamics around an electrically charged Kiselev black hole embedded in quintessence}

\vspace{48pt}

\textbf{Vitalie Lungu, }\footnote{Corresponding author e-mail: \texttt{vitalie.lungu@student.uaic.ro }}
\textbf{Marina-Aura Dariescu, }\footnote{E-mail: \texttt{marina@uaic.ro}}
\textbf{ Cristian Stelea,}\footnote{E-mail: \texttt{cristian.stelea@uaic.ro}}

\vspace*{0.2cm}

\textit{$^{1,2}$ Faculty of Physics, ``Alexandru Ioan Cuza" University of Iasi}\\[0pt]
\textit{11 Bd. Carol I, Iasi, 700506, Romania}\\[.5em]

\textit{$^3$ Department of Exact and Natural Sciences, Institute of Interdisciplinary Research,}\\[0pt]
\textit{``Alexandru Ioan Cuza" University of Iasi}\\[0pt]
\textit{11 Bd. Carol I, Iasi, 700506, Romania}\\[.5em]

\end{center}

\vspace{30pt}

\begin{abstract}
We introduce and study a new solution describing a static, spherically symmetric and electrically charged black hole embedded in a charged quintessence fluid, which corresponds to an electric generalization of the Kiselev geometry. We derive the effective potential and analyze the various types of orbits followed by charged particles. A special attention is given to circular orbits and their stability. We found that for uncharged particles the periapsis shifts for bounded orbits is always prograde. However, for charged test particles the periapsis shifts can become retrograde in some cases.
\end{abstract}

\vspace{32pt}

\setcounter{footnote}{0}

\newpage

\section{Introduction}

Black holes have attracted sustained attention over the past several decades owing to their fundamental role in the framework of general relativity. Recent advances in gravitational-wave astronomy \cite{LIGOScientific:2016aoc, LIGOScientific:2020ibl} and very-long-baseline interferometry \cite{EventHorizonTelescope:2019dse} have inaugurated a new era in black hole physics. These observational techniques provide unprecedented opportunities to test theoretical predictions and to impose constrains on the black holes parameters. The uniqueness of black holes is encapsulated in the so-called “no-hair” theorem, which states that they are completely characterized by only three parameters: mass, angular momentum and electric charge. However, in realistic astrophysical settings, black holes are not isolated objects. Rather, they are embedded in complex environments consisting of electromagnetic fields, plasma, dark matter and dark energy. The presence of a dominant dark energy with negative pressure was imposed by recent supernova observations indicating an accelerated expansion of the Universe \cite{SupernovaSearchTeam:1998fmf} and one of the candidates for the dark energy is the quintessence (see \cite{Tsujikawa:2013fta} for a review).

Black holes embedded in anisotropic fluids were first described by Brown and Husain in \cite{Brown:1997jv} as black holes with short hair. They correspond to spherically symmetric black hole solutions for Einstein gravity coupled to anisotropic fluids. In the regime where the parameter $k<\frac{1}{2}$ the same fluid solution from  \cite{Brown:1997jv} was later rediscovered by Kiselev \cite{Kiselev:2002dx} as the so-called black hole solution surrounded by a ``quintessence'' fluid. 
This anisotropic fluid is satisfying the equation of state $p=\rho w$,\footnote{Here $p$ is an isotropic pressure.} where the parameter  $w$  has to be in the range $w \in (-1, 1/3)$ in order to cause the accelerated expansion of the Universe. Following Kiselev's seminal work, a lot of physicists have used the Kiselev solution to study how the black hole’s spacetime is modified by the surrounding quintessence fluid (see for example \cite{Azreg-Ainou:2017obt, Konoplya:2019sns, Zeng:2020vsj, Abdujabbarov:2015pqp}). 
More recently, the Kiselev geometry has been reinterpreted as an exact solution in the context of nonlinear electrodynamics \cite{Dariescu:2022kof} and the motion of charged particles in this background was studied in \cite{Dariescu:2023twk}. 

In the present work, we are going to consider a generalization of this geometry, by presenting a new exact solution that describes an electrically charged Kiselev black hole surrounded by a charged anisotropic quintessence fluid. This solution is to be contrasted to the solutions presented in \cite{Kiselev:2002dx},  \cite{Jeong:2023hom} (see also references therein) that describe charged black holes surrounded by ``quintessence''. In our case the anisotropic fluid is electrically charged and it contributes as a source to the Maxwell equations for the electromagnetic field, unlike the previous solutions known in literature. This solution can be obtained using the results from \cite{Stelea:2018elx} - \cite{Stelea:2018cgm}.

Nowadays, it is widely believed that at the center of each large galaxy lies a supermassive black hole \cite{Kormendy:2013dxa}. This is what happens in our Milky Way galaxy as well \cite{Genzel2010}. The supermassive black hole in the center of our galaxy was named Sagittarius A$^*$  (Sgr A$^*$) and, in 2022, the first image of Sgr A* was released by the Event Horizon Telescope Collaboration \cite{EventHorizonTelescope:2022wkp} - \cite{EventHorizonTelescope:2024rju}. While the supermassive black hole is usually modeled using a vacuum Kerr geometry, which is appropriate to describe a rotating black hole, the true nature of the supermassive compact object Sgr A$^*$ is still open to debate \cite{Vagnozzi:2022moj}. One should note that Sgr A$^*$ is surrounded by a disk of young stars and inside the inner radius of this disk there is the so-called the S-cluster of young stars \cite{Ghez:2003rt} and there are also strong magnetic fields in its vicinity \cite{Eatough:2013nva}. Therefore, one of the most effective ways to probe into the nature of Sgr A$^*$ is to trace the orbits of the S-cluster stars. These stars can be considered as massive particles which follow bound timelike orbits around the central object. For example, the gravitational redshift of the S2 star \cite{Gillessen:2008qv} and the relativistic precession of the S2 star orbit have been recently reported in \cite{GRAVITY:2020gka}.

In this general context, in the present work, we analyze the timelike geodesics for charged particles, which reveal the structure of stable and unstable circular orbits. Future observations of M87*, Sgr A*, and other supermassive black holes promise ever tighter tests of gravity in the strong-field regime. Any deviations from the Kerr or Schwarzschild predictions for instance, due to exotic environments could be constrained by shadow measurements and light-deflection data. Our results can be compared to similar investigations on particle's motion in Kiselev, Reissner--Nordstr\"om and other important spacetimes (see for example \cite{Fernando:2012ue}- \cite{Garnier:2025jnp}).

When dealing with quasi-circular trajectories, one of the most important general relativistic effects in orbital motions around compact objects is the periapsis shift. The most famous example of this phenomenon was used by Einstein to explain the perihelion shift of Mercury in its motion around the Sun: in this case the elliptic orbit of Mercury rotates in the same direction as its orbital evolution around the Sun. One speaks about a prograde precession of Mercury's orbit in this case. However, even if the general relativistic effects imply the existence of a prograde periapsis shift in the motion of a star around a massive compact object, there are also cases in which this periapsis shift can become retrograde due to various reasons. One such reason can be the presence of local matter density around the compact object \cite{Harada:2022uae} - \cite{Igata:2022nkt}. Retrograde periapsis precessions can also occur in spacetime that contain naked singularities \cite{Bini:2005dy} - \cite{Ota:2021mub}. In fact, the relativistic periapsis precession of the S2 star can be either retrograde or prograde depending on the amount of dark matter enclosed within its orbit (see also \cite{Arguelles:2021jtk}).
In this paper we further propose to extend the study of the periapsis shifts in the case of the  electrically charged Kiselev black holes. We found that for uncharged test particles there is no retrograde periapsis shift. However, for charged test particles there are values of the parameters for which the retrograde periapsis shift is possible.

The paper is organized as it follows: In section $2$ we introduce the metric describing an electrically charged black hole embedded in an electrically charged quintessence fluid and briefly discuss its horizons structure. In section $3$ we discuss the motion of timelike charged particles in this geometry. The effective potential derived in subsection $3.1$ allows us to analyze the type of orbits followed by charged particles. A special attention is given to circular orbits which are widely discussed in subsection $3.3$. The stability of circular orbits in terms of the Lyapunov exponent is analyzed in the section $3.4$, for both stable and unstable circular orbits. Subsection $3.7$ deals with the special case of charged particles with zero angular momentum.
In section $4$, using the Hamiltonian method, we study the periapsis shift, pointing out the possibility of the retrograde motion for charged particles moving around an electrically charged Kiselev black hole. The paper ends with concluding remarks in section $5$.

\section{The Electrically charged Kiselev black hole}

The Kiselev geometry is described by the following static four-dimensional line-element  \cite{Kiselev:2002dx}:
\beqs
ds^2&=&= - f dt^2+\frac{dr^2}{f} + r^2 ( d \theta^2 + \sin^2 \theta d \varphi^2)
\label{kiselev} \; ,
\eeqs
where
\begin{equation}
f= 1 -\frac{2M}{r} - \frac{k}{r^{3 w +1}}.
\label{metric}
\end{equation}
Here $w$ is the equation of state parameter and $k$ is a positive quintessence parameter, which is related to the fluid quintessence energy density:
\beqs
\rho^0 &=& - \frac{3kw}{8\pi r^{3(w+1)}},
\eeqs
while the components of the anisotropic pressures can be written as $p_r^0=-\rho^0$ and the tangential pressures $p_t^0$ are given by:
\beqs
p_t^0&\equiv&p_{\theta}^0=p_{\varphi}^0=-\frac{3(3w+1)kw}{16 \pi r^{3(w+1)}}.
\eeqs
In order to have an accelerated expansion, the equation of state parameter $w$ should belong to the interval $w \in [ -1 , -1/3]$. 

Concerning the values of the quintessence parameter $k$, several constraints have been obtained by confronting theoretical predictions with the observational data reported by the Event Horizon Telescope (EHT) Collaboration for the supermassive black holes SgrA* and M87* \cite{EventHorizonTelescope:2019dse}. 
For instance, in \cite{Atamurotov:2022nim}, the authors assumed that these astrophysical objects can be described as Kerr supermassive black holes surrounded by quintessence. By analyzing the shadow observables, they derived upper bounds for the quintessence parameter given by $kM = 0.011$ for the SgrA* black hole and $kM = 0.0358$ for the M87* black hole. 
The constraint obtained for SgrA* is consistent with the recent estimate derived for a Frolov black hole surrounded by quintessence, which leads to the bound $kM \leq 0.012$ \cite{Gohain:2024piy}.

Using now the results from \cite{Stelea:2018elx} - \cite{Stelea:2018cgm} we consider the electrically charged black hole embedded in electrically charged ``quintessence'' fluid as being described by the line element \cite{Stelea:2018elx}:
\begin{equation}
ds^2= - \frac{f}{\Lambda^2} dt^2 + \Lambda^2 \left[ f^{-1} dr^2 + r^2 \left( d \theta ^2 + \sin^2 \theta d \varphi^2 \right) \right]  ,
\label{metelec}
\end{equation}
where 
\begin{equation}
\Lambda =\frac{1-U^2 f}{C},
\label{lambda}
\end{equation}
while $0\leq U<1$ and $C$ are constants. For an asymptotically flat exterior solution, the constant $C$ must be taken as $C=1-U^2$. \footnote{In our case, since the Kiselev geometry is not asymptotically flat, one can set the value of $C$ to $1$ without losing generality.} The parameter $U$ is generally connected to the electric charge of the black hole. If $U=0$ one recovers the uncharged Kiselev black geometry. The equation of state parameter $w$ must be in the range $w \in [-1 , -1/3]$ in order to induce a cosmological acceleration effect in the final solution as well. In our case, there is an additional electric field which is generated by the electric potential\footnote{Note that $\Lambda$ is a function of the radial coordinate $r$, not to be confused to the cosmological constant.}:
\begin{equation}
A_t \, = - \, \frac{Uf}{\Lambda},
\label{A0}
\end{equation}
while the energy density of the electrically charged quintessence fluid is given by $\rho=\frac{\rho^0}{\Lambda^2}+\rho_e$ and the anisotropic pressures are given by $p_r=\frac{p_r^0}{\Lambda^2}$, $p_t=\frac{p_t^0}{\Lambda^2}$. The energy density $\rho$ contains a contribution $\rho_e=\frac{4p_t}{C}\frac{U^2f}{\Lambda}$ due to the charged ``quintessence'' fluid, while the Maxwell field equations are now modified and contain a source term with current $J_{\mu}=-\frac{4p_t}{C}\frac{Uf}{\Lambda^2}\delta_{\mu}^t$ (for more details see   \cite{Stelea:2018elx}, \cite{Stelea:2018shx}).

There is a physical motivation behind the ansatz used for the electrical potential in (\ref{A0}). Note that with this expression there is a quadratic functional dependence of the $tt$ component of the metric, $g_{tt}$, and the electric potential in the form $g_{tt}=-A_t^2+\frac{1-U^2}{U}A_t$. This Weyl-type relation was first derived by Weyl in the pioneering work \cite{Weyl:1917gp} on static Einstein-Maxwell systems. Later, Gautreau and Hoffmann showed in \cite{GH}  (see also \cite{Lemos:2009mr}) that all the possible fluid sources that could lead to such quadratic functional relations of the Weyl-type should be described by an electric potential and electric density charge $\rho_e$ as described above.

Then the system of Einstein-Maxwell-fluid equations of motion:
\beqs
G_{\mu\nu}&=&8\pi T_{\mu\nu}+8\pi T_{\mu\nu}^{em},\nonumber\\
F^{\mu\nu}_{~~;\nu}&=&4\pi J^{\mu}
\eeqs
is satisfied if one considers the stress-energy of the fluid as being given by:
\beqs
T_{\mu\nu}&=&(\rho+\rho_e+p_t)u_{\mu}u_{\nu}+p_t g_{\mu\nu} +(p_r-p_t)\chi_{\mu}\chi_{\nu},
\label{Tfluid}
\eeqs
while the stress-energy tensor of the electromagnetic field can be written in the usual form:
\beqs
T^{em}_{\mu\nu}&=&\frac{1}{4\pi}\left(F_{\mu\gamma}F_{\nu}^{~\gamma}-\frac{1}{4}g_{\mu\nu}F_{\gamma\delta}F^{\gamma\delta}\right).
\eeqs
Here $u^{\mu}=\frac{\Lambda}{\sqrt{f}}\delta^{\mu}_t$ is the $4$-velocity of the fluid, normalized such that $u^{\mu}u_{\mu}=-1$, while $\chi^{\mu}=\frac{\sqrt{f}}{\Lambda}\delta^{\mu}_r$ is the unit vector in the radial direction. 

\subsection{The horizons structure}
The spacetime horizons for the electrically charged Kiselev black hole are again given by the solution of the equation $f=0$, the same as in the original uncharged Kiselev geometry. Depending on the values of $k$ and $w$, there can be up to two horizons. The inner one corresponds to the black hole horizon, while the outer one is the ``cosmological'' horizon, common to the Kiselev geometry. As an example, for $w=-2/3$, the metric function (\ref{metric}) becomes:
\begin{equation}
f=1-\frac{2M}{r}-kr,
\label{metric23}
\end{equation}
and the horizons have the simple expressions:
\begin{equation}
r_{\pm}=\frac{1\pm\sqrt{1-8kM}}{2k},
\label{rpm}
\end{equation}
where $r_-$ corresponds to the black hole horizon, while $r_+$ is the cosmological one. For a fixed $w$, the value of $r_-$ increases with $k$. 
In order to find the values of $k \equiv k_*$ for which the two horizons coincide, i.e. $r_-=r_+ \equiv r_*$, one has to solve the system $f=f'=0$ which has the solutions
\begin{equation}
r_*=\frac{6Mw}{1+3w},  \qquad k_*=-\frac{2M (6Mw)^{3w}}{(1+3w)^{1+3w}}.
\label{rw}
\end{equation}

\section{Electrically charged particles in presence of an electric Kiselev black hole}

Let us describe now the motion of a charged test particle in the background of an electrically charged Kiselev black hole.

\subsection{The effective potential}

The motion of a charged particle (with unit mass) can be described using the Hamiltonian \cite{Wald}:
\[
H = \frac{1}{2} g^{\mu\nu} (\pi_{\mu}-\varepsilon A_{\mu})(\pi_{\nu}-\varepsilon A_{\nu}), 
\]
where $\pi_{\mu}=m\dot{x}_{\mu}+qA_{\mu}$  is the canonical momentum of the charged particle, $\varepsilon=q/m$ represents the specific charge of the particle and $A_{\mu}$ is the electromagnetic potential. 

Due to symmetry of the spacetime, one can find the conserved quantities, namely the energy $E$ and the angular momentum $L$ which are given by the relations:
\begin{equation}
E=-\pi_t=\frac{f}{\Lambda}\left(\frac{\dot{t}}{\Lambda}+\varepsilon U\right),
\label{E}
\end{equation}
\begin{equation}
L=\pi_{\varphi}=\Lambda^2 r^2 \sin^2 \theta \dot{\varphi},
\label{L}
\end{equation}
while the other components of the momentum are: $\pi_r=\frac{\Lambda^2}{f}\dot{r}$ and $\pi_{\theta}=\Lambda^2 r^2 \dot{\theta}$. In view of the above relations, the Hamiltonian can be written as
\begin{equation}
H=\frac{f}{2 \Lambda^2}\pi_r^2+\frac{1}{2\Lambda^2r^2}\pi_{\theta}^2+\frac{1}{2}\left[\frac{L^2}{\Lambda^2 r^2\sin^2\theta}-\frac{\Lambda^2}{f}(E+\varepsilon A_t)^2\right].
\label{H1}
\end{equation}
Without loss of generality, due to the spherical symmetry of the system, one can restrict the motion to the equatorial plane, $\theta=\pi/2$. One may separate the potential part of the Hamiltonian (\ref{H1}):
\begin{equation}
H_{pot} = \frac{1}{2} \left[ - \frac{\Lambda^2 }{f} \left( E - \frac{\varepsilon U f}{\Lambda} \right)^2 + \frac{L^2}{r^2 \Lambda^2}  \right].
\label{HpotEl}
\end{equation}

The total Hamiltonian $H$ is a constant, i.e. $H=-m^2/2$ and  by setting $m=1$, one obtains the relation:
\begin{equation}
\dot{r}^2+\left[\frac{f L^2}{\Lambda^4 r^2}-( E + \varepsilon A_t)^2\right]=-\frac{f}{\Lambda^2},
\label{firstintegral}
\end{equation}
which can be further written in the form:
\begin{equation}
\dot{r}^2=( E - V_-)(E-V_+),
\end{equation}
where $V_-$ and $V_+$ are the two roots of the equation $\dot{r}^2=0$. In the followings, we shall restrict to the case corresponding to $V_+$ which is a positive quantity and select it as the effective potential. This has the expression:
\begin{equation}
V_{eff}=\frac{\varepsilon Uf}{\Lambda}+\sqrt{\frac{f}{\Lambda^2}\left(1+\frac{L^2}{\Lambda^2r^2}\right)},
\label{potential}
\end{equation}
and is represented in the following figures:

\begin{figure}[H]
    \centering
    \begin{subfigure}{0.49\textwidth}
        \centering
        \includegraphics[scale=0.5, trim=2cm 10cm 0cm 1cm, clip]{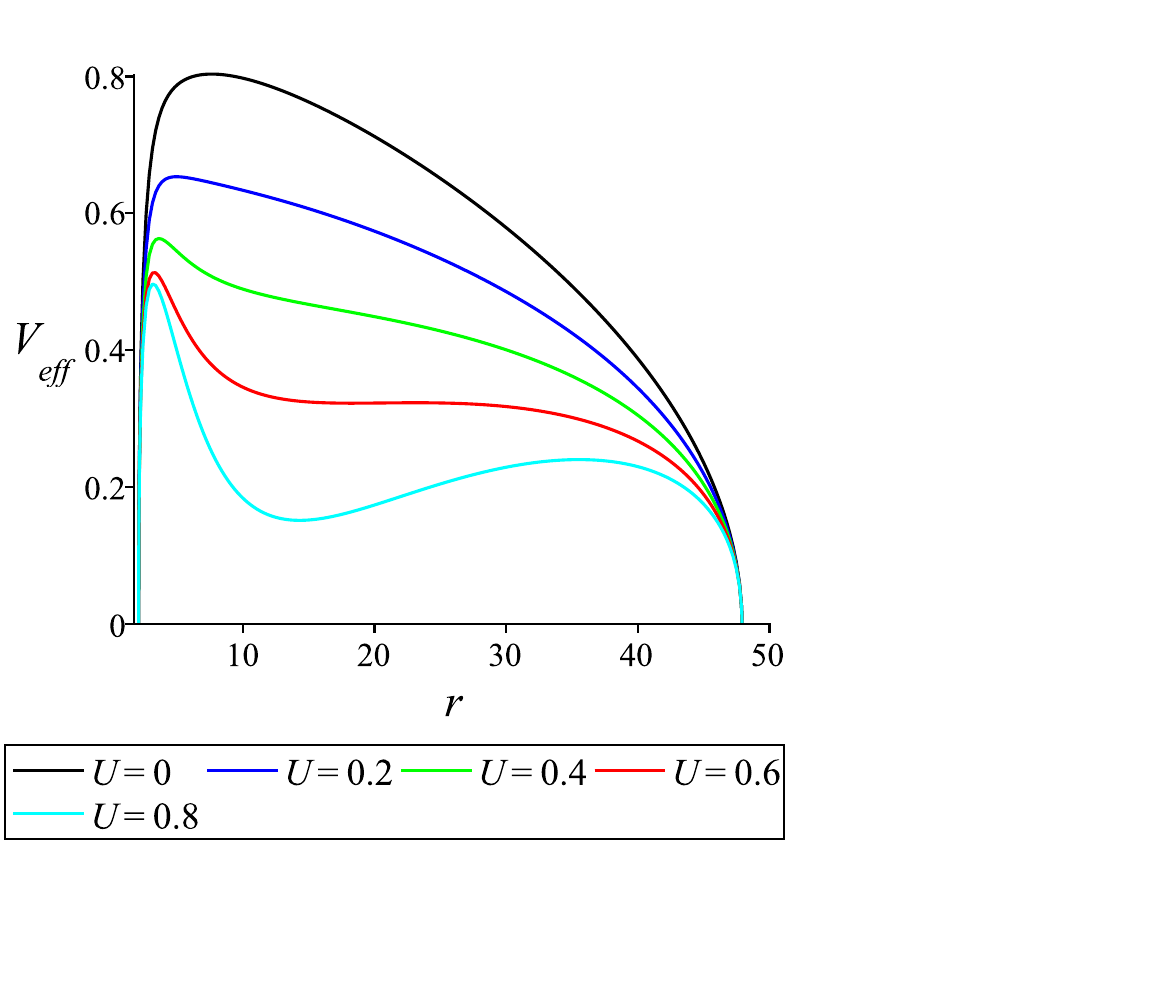}
    \end{subfigure}
    \hfill
    \begin{subfigure}{0.49\textwidth}
        \centering
        \includegraphics[scale=0.5, trim=0cm 10cm 0cm 1cm, clip]{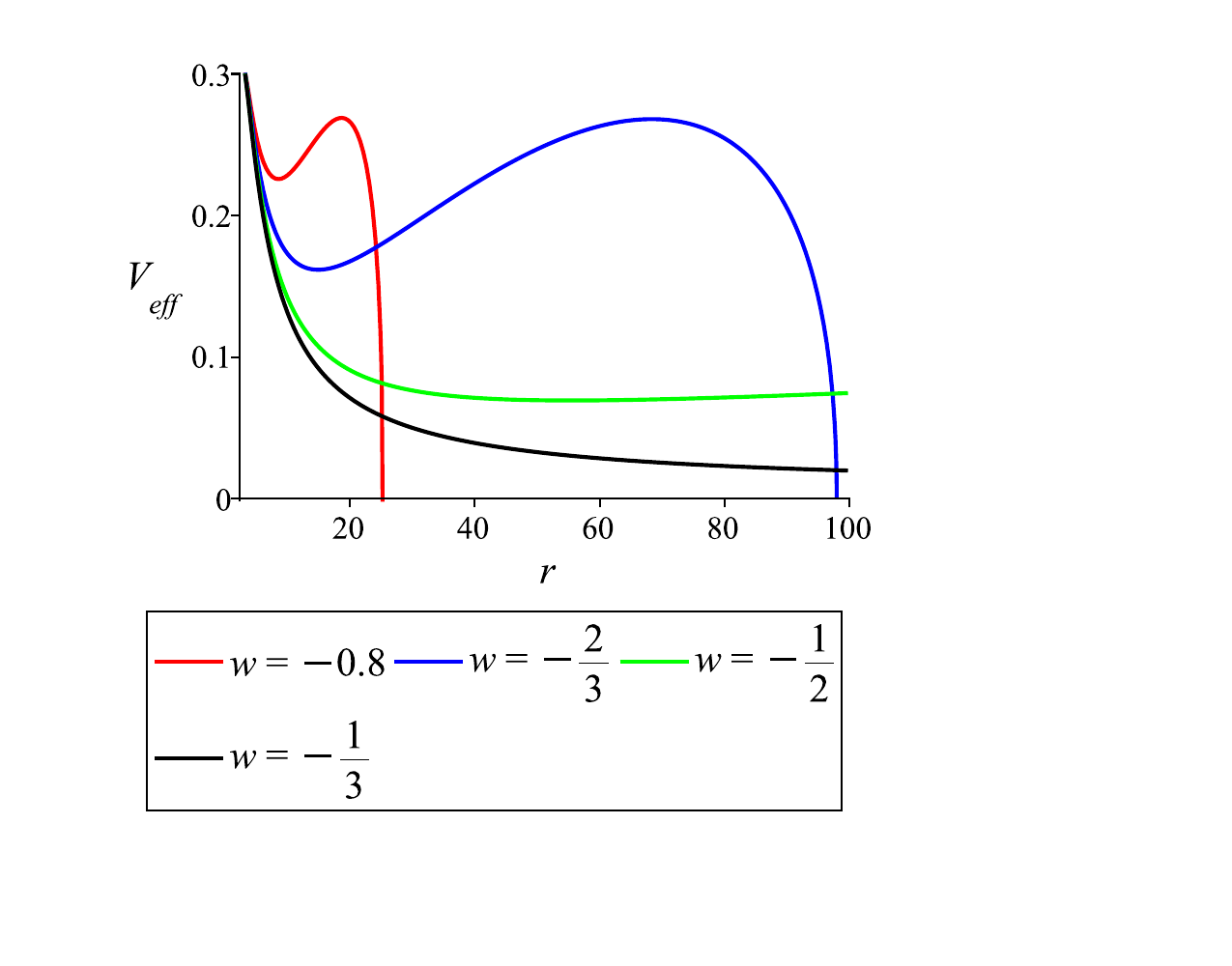}
    \end{subfigure}
    \caption{{\it{Left panel.}} The effective potential (\ref{potential}) for different values of $U$. The values of the other parameters are: $M=1$, $w=-2/3$, $k=0.02$, $\varepsilon=-3/2$, $L=\sqrt{6}$. {\it{Right panel}}. The effective potential (\ref{potential}) for different values of $w$. The values of the other parameters are: $M=1$, $U=0.5$, $k=0.01$, $\varepsilon=-2$ and $L=1$. }
    \label{fig:potU}
\end{figure}

\begin{figure}[H]
    \centering
    \begin{subfigure}{0.49\textwidth}
        \centering
        \includegraphics[scale=0.5, trim=2cm 10cm 0cm 1cm, clip]{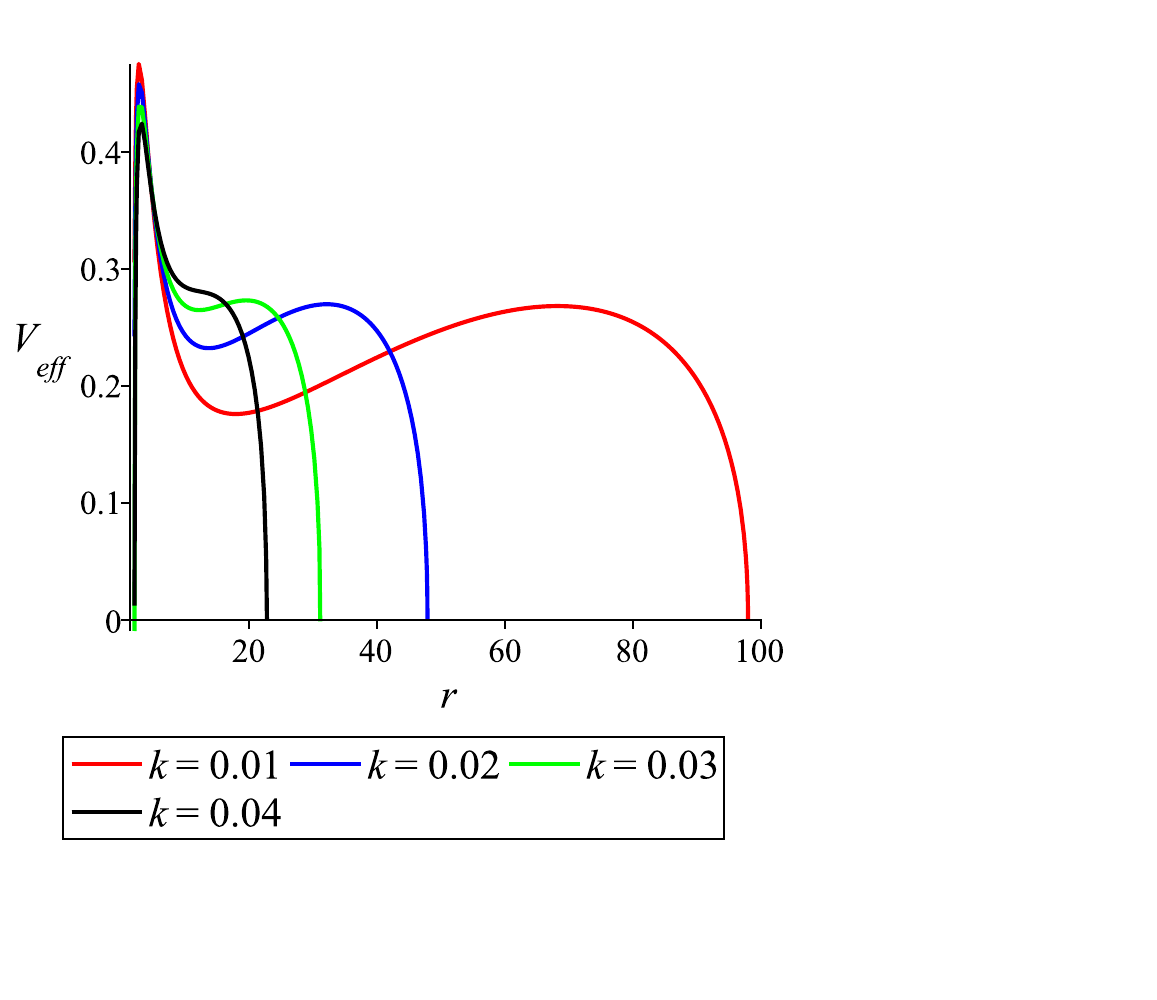}
    \end{subfigure}
    \hfill
    \begin{subfigure}{0.49\textwidth}
        \centering
        \includegraphics[scale=0.5, trim=0cm 10cm 0cm 1cm, clip]{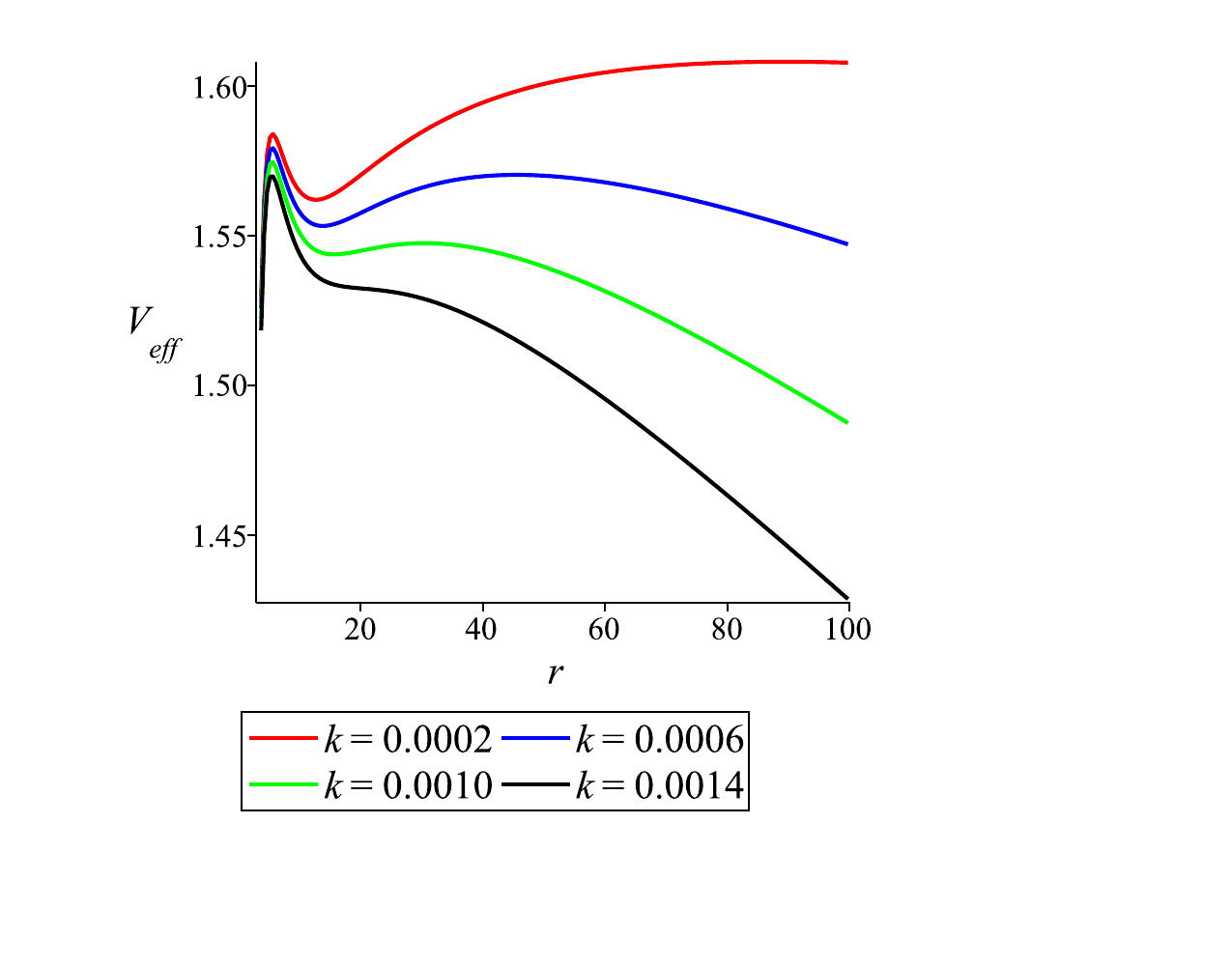}
    \end{subfigure}
    \caption{The effective potential (\ref{potential}) for different values of $k$. The values of the parameters are: $M=1$, $w=-2/3$, $U=0.5$, $\varepsilon=-2$, $L=2.5$ ({\it left panel}) and $M=1$, $w=-2/3$, $U=0.5$, $\varepsilon=0.5$, $L=5$ ({\it right panel}).}
    \label{fig:potk}
\end{figure}

The effective potential (\ref{potential}) generally depends on three parameters $w$, $k$ and $U$ characterizing the anisotropic fluid and the electric charge (we set the black hole mass to $M=1$). In Figures \ref{fig:potU} and \ref{fig:potk} one can see its radial shape for different values of these parameters. However, for simplicity, we will mainly use the value $w=-2/3$, for which the metric function $f(r)$ takes the particularly simple form (\ref{metric23}).
By inspecting the figures \ref{fig:potU} and \ref{fig:potk}, one may notice that the effective potential (\ref{potential}) vanishes on the horizons and its shape is determined by the values of the parameters $w$, $k$ and $U$. There is always a maximum situated just outside the event horizon $r_-$. Also, for large values of $U$ or $|w|$, the effective potential has a second maximum close to the cosmological horizon and a minimum value between them. Thus, particles with suitable energies can be trapped in bound orbits.

As it can be noticed in the figure \ref{fig:potk}, the effective potential (\ref{potential}) is also influenced by the quintessence parameter $k$. As $k$ increases, the cosmological horizon and the second maximum of the potential move to smaller radial distances and the range $r_- \leq r \leq r_+$ shrinks. There is a critical value of $k$ above which the second maximum vanishes and no bound orbits are allowed.

\subsection{The types of orbits of the charged test particles}

\begin{figure}[H]
 \centering
    \begin{subfigure}{0.49\textwidth}
        \centering
        \includegraphics[scale=0.5, trim=2cm 12cm 5cm 1cm, clip]{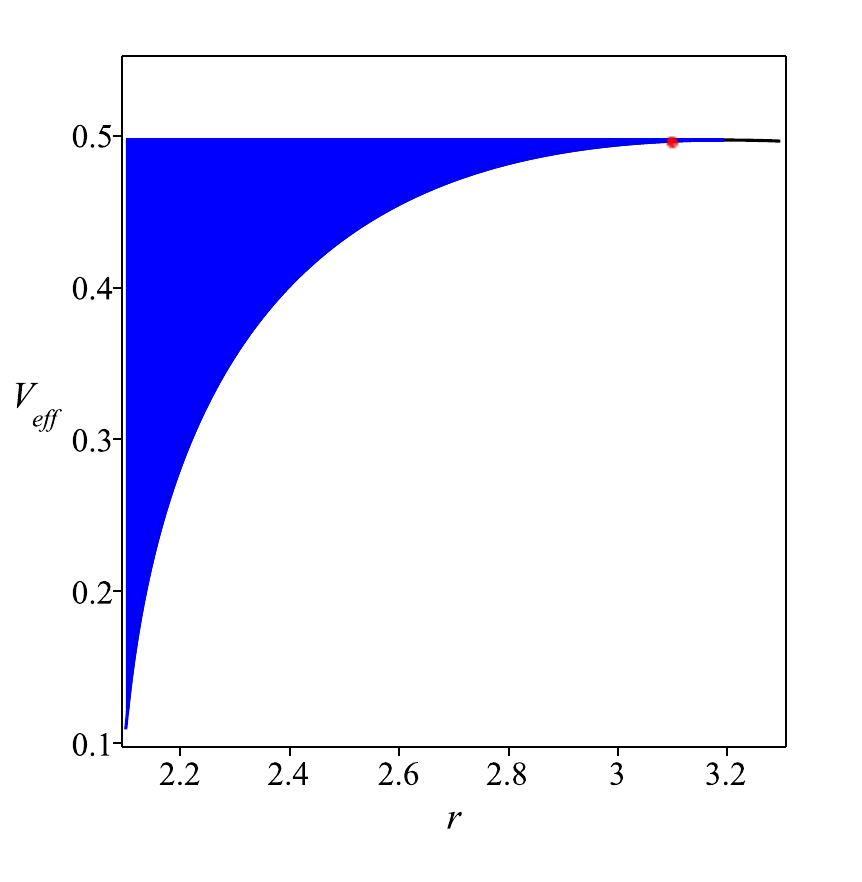}
    \end{subfigure}
    \hfill
    \begin{subfigure}{0.49\textwidth}
        \centering
        \includegraphics[scale=0.5, trim=0cm 12cm 5cm 1cm, clip]{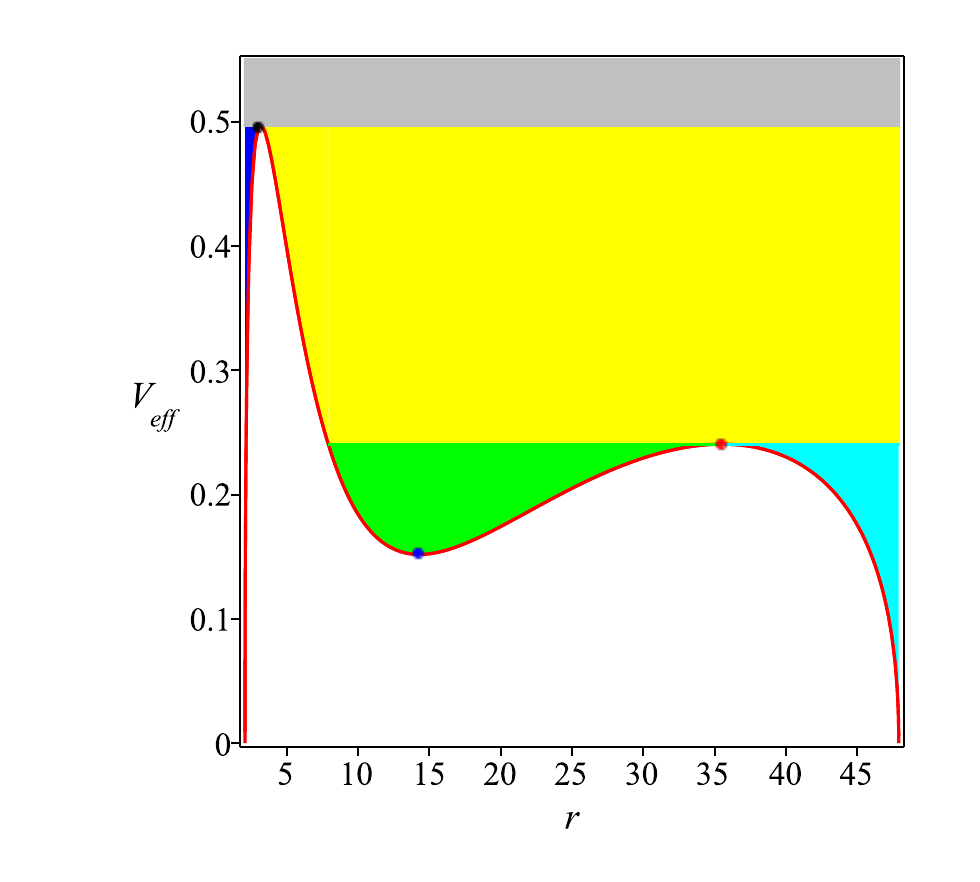}
    \end{subfigure}
        \begin{tikzpicture}[overlay]
\node at (5, 4){};
\draw[->, thick] ( -11, 5) -- (-6.9,5.9);
\end{tikzpicture}
    \caption{Regions corresponding to $E \geq V_{eff}$,  where the particle motion is allowed. The values of the parameters are: $M=1$, $U=0.8$, $w=-2/3$, $k=0.02$, $\varepsilon=-3/2$, $L=\sqrt{6}$.}
    \label{fig:regions}
\end{figure}

In order to analyse the types of orbits, we start with the second order differential equation describing the radial motion which can be obtained from (\ref{firstintegral})
as being:
\begin{equation}
\ddot{r}=\left(\frac{f'}{2f}-\frac{\Lambda'}{\Lambda}\right)\left[\dot{r}^2-\left(E-\frac{\varepsilon U f}{\Lambda}\right)^2\right]+\frac{f(\Lambda+\Lambda'r) L^2}{\Lambda^5 r^3}+\frac{\varepsilon U (f\Lambda'-f'\Lambda^2)}{\Lambda^3}\left(E-\frac{\varepsilon U f}{\Lambda}\right)
\label{eqmotion}
\end{equation} 
The orbits can be obtained by numerically integrating the above equation
for a particle moving in the potential (\ref{potential}). In the figure \ref{fig:regions},  the allowed regions correspond to the physical relation $E \geq V_{eff}$ and are represented in coloured areas. The critical points are highlighted by black, blue and red dots. 

\begin{figure}[H]
\captionsetup[subfigure]{justification=centering}
\centering
  \begin{subfigure}{.42\textwidth}
  \centering
    \includegraphics[width=0.9\linewidth]{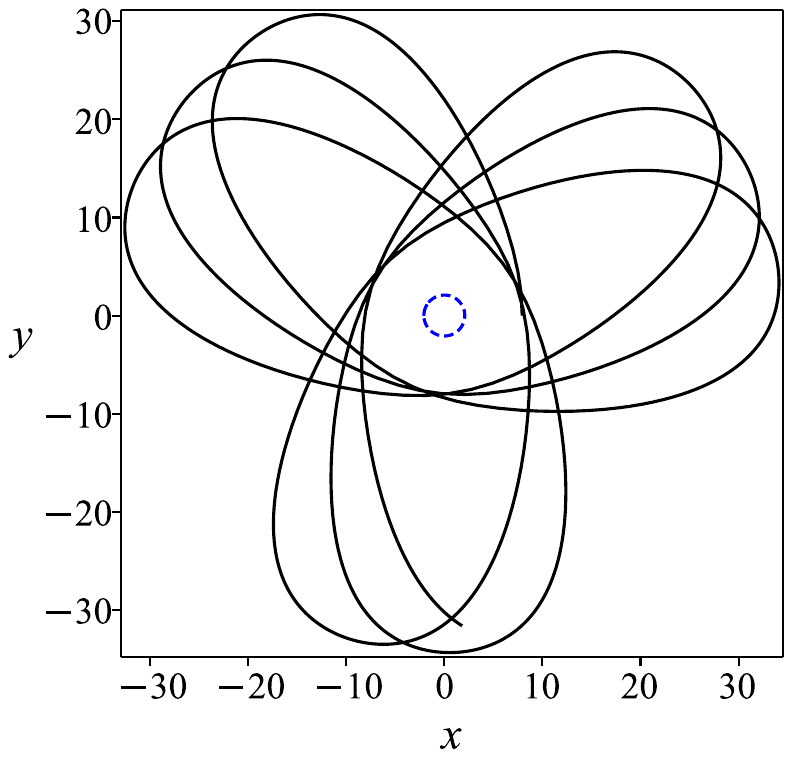}
    \caption*{\hspace{-10mm} a) $E=0.240$}
  \end{subfigure}\hspace{-30mm}
  \label{a}
  \begin{subfigure}{.42\textwidth}
  \centering
    \includegraphics[width=0.9\linewidth]{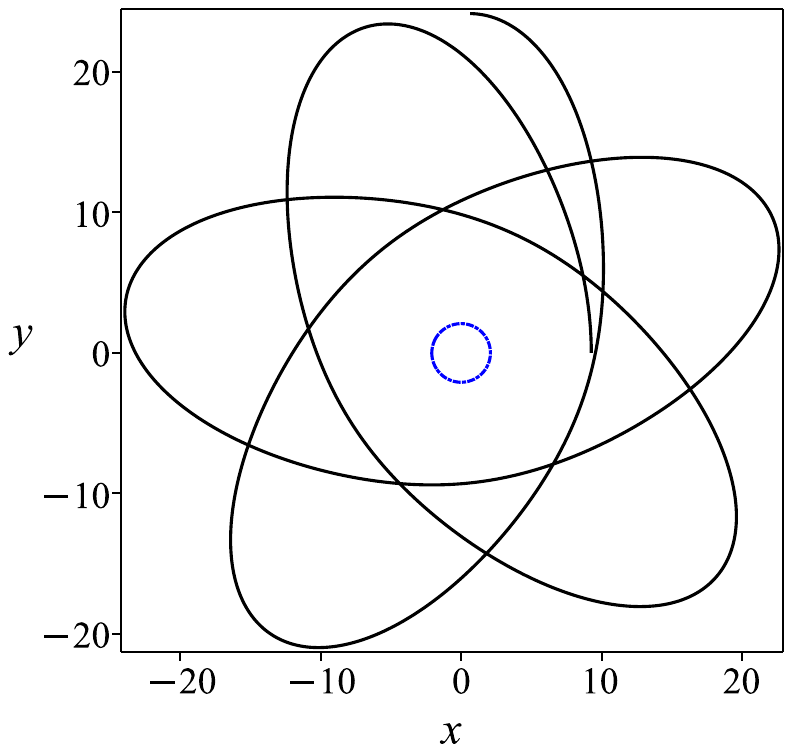}
    \caption*{\hspace{-10mm} b) $E=0.200$}
  \end{subfigure}\hspace{-30mm}
   \begin{subfigure}{.42\textwidth}
   \centering
    \includegraphics[width=0.9\linewidth]{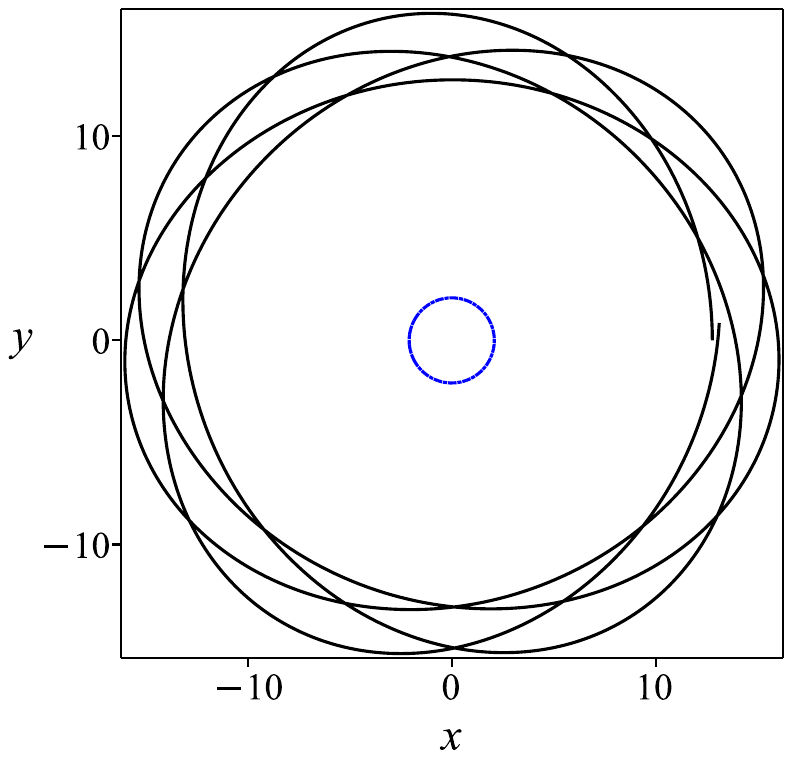}
    \caption*{\hspace{-10mm} c)$E=0.155$}
  \end{subfigure}\hspace{-30mm}
  \caption{Bound orbits of charged particles moving in the potential represented in figure \ref{fig:regions}. The blue circle represents the black hole horizon $r_-$.}
  \label{fig:orbitselectric}
\end{figure}

\begin{figure}[H]
\captionsetup[subfigure]{justification=centering}
\centering
  \begin{subfigure}{.42\textwidth}
  \centering
    \includegraphics[width=0.9\linewidth]{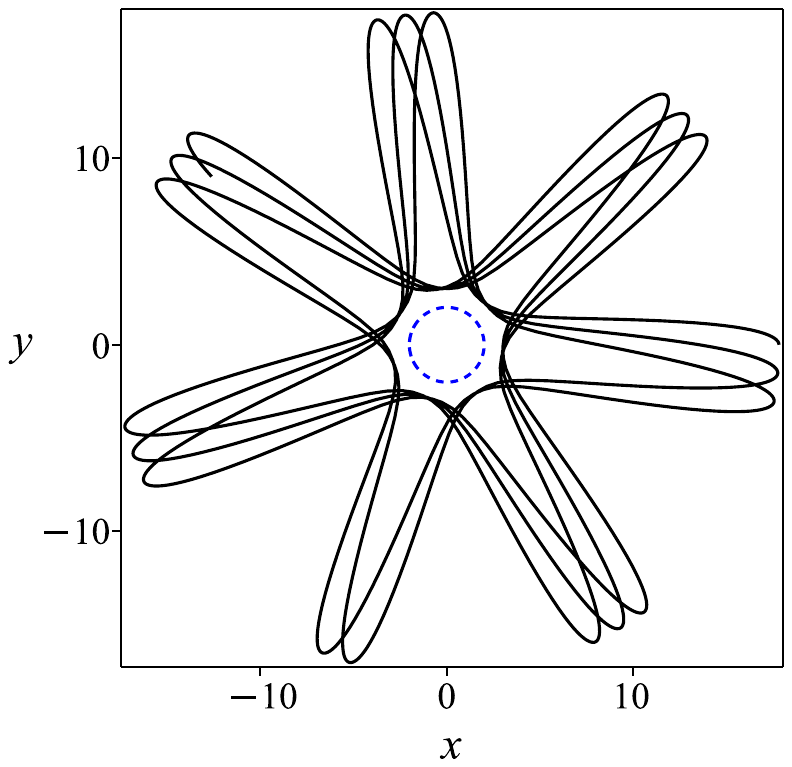}
    \caption*{\hspace{-10mm} a) $E=0.20$}
  \end{subfigure}\hspace{-30mm}
  \label{a}
  \begin{subfigure}{.42\textwidth}
  \centering
    \includegraphics[width=0.9\linewidth]{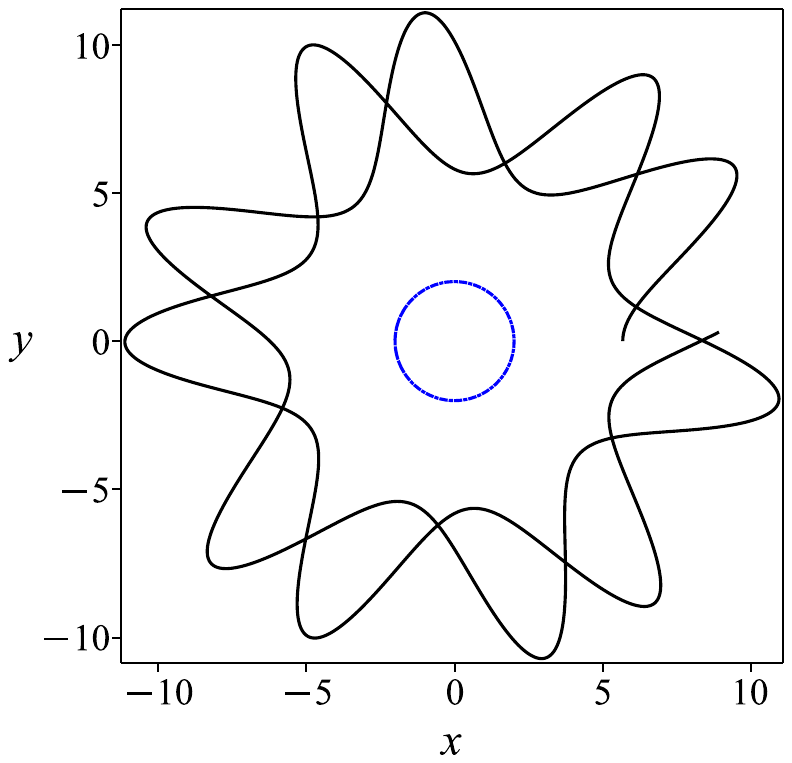}
    \caption*{\hspace{-10mm} b) $E=0.05$}
  \end{subfigure}\hspace{-30mm}
   \begin{subfigure}{.42\textwidth}
   \centering
    \includegraphics[width=0.9\linewidth]{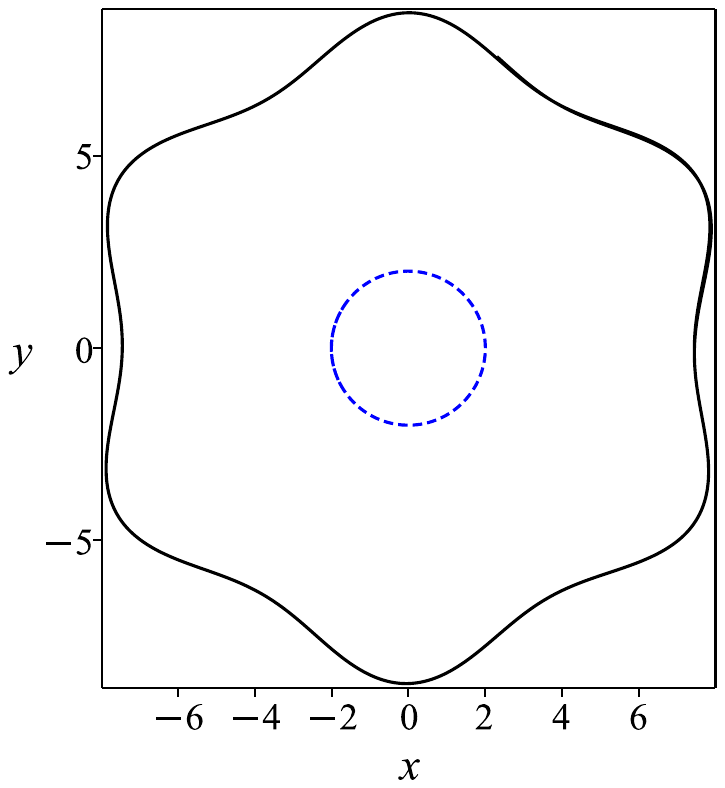}
    \caption*{\hspace{-10mm} c)$E=0.02$}
  \end{subfigure}\hspace{-30mm}
  \caption{Bound orbits of charged particles. The values of parameters are $M=1$, $U=0.5$, $k=0.002$, $w=-1$,  $\varepsilon=-2.5$, $L=0.7$. The blue circle represents the black hole horizon $r_-$.}
  \label{fig:orbitselectric1}
\end{figure}

\begin{figure}[H]
\captionsetup[subfigure]{justification=centering}
\centering
  \begin{subfigure}{.42\textwidth}
  \centering
    \includegraphics[width=0.9\linewidth]{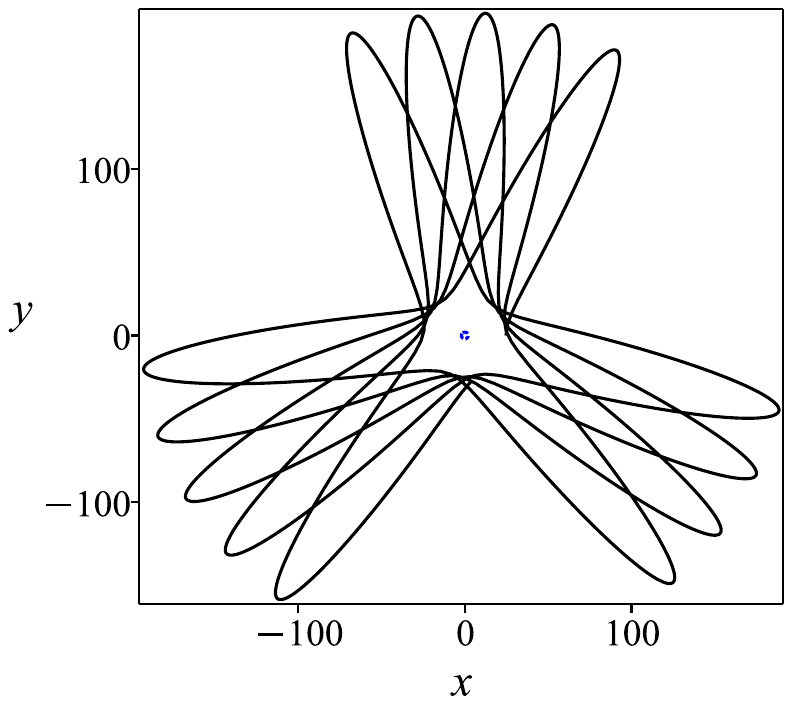}
    \caption*{\hspace{-10mm} a) $E=0.280$}
  \end{subfigure}\hspace{-30mm}
  \label{a}
  \begin{subfigure}{.42\textwidth}
  \centering
    \includegraphics[width=0.9\linewidth]{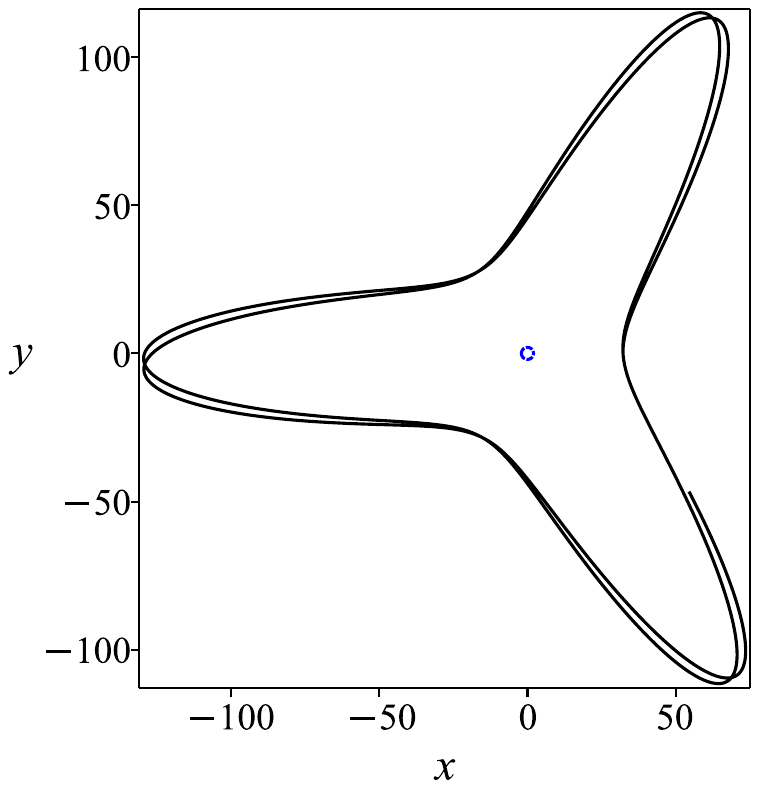}
    \caption*{\hspace{-10mm} b) $E=0.275$}
  \end{subfigure}\hspace{-30mm}
   \begin{subfigure}{.42\textwidth}
   \centering
    \includegraphics[width=0.9\linewidth]{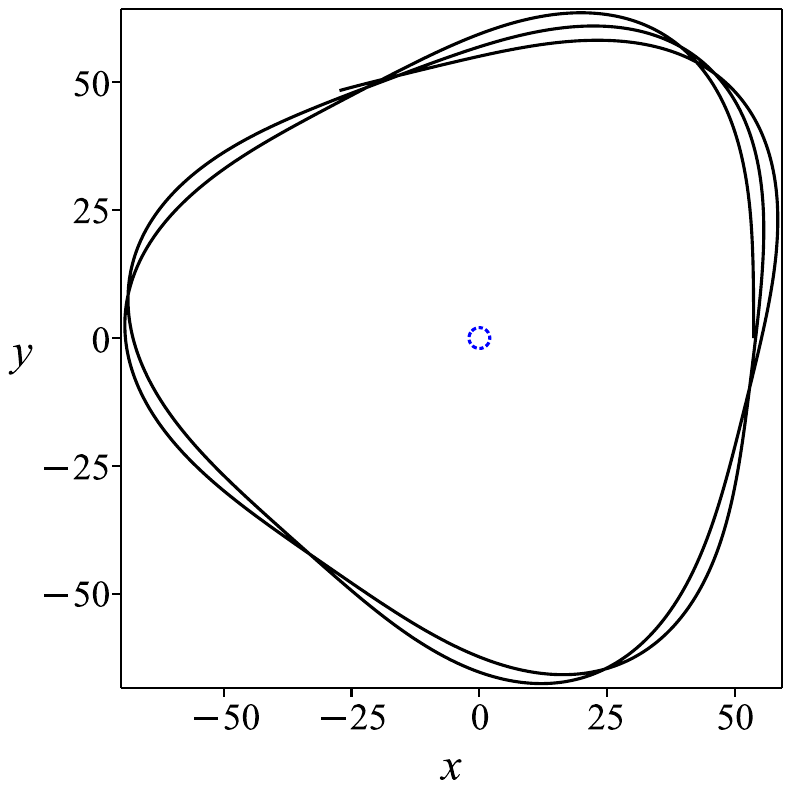}
    \caption*{\hspace{-10mm} c)$E=0.271$}
  \end{subfigure}\hspace{-30mm}
  \caption{Bound orbits of charged particles. The values of parameters are $M=1$, $U=0.4$, $k=0.01$, $w=-1/2$,  $\varepsilon=-2$, $L=2$.The blue circle represents the black hole horizon $r_-$.}
  \label{fig:orbitselectric2}
\end{figure}

As it can be noticed, the potential represented in the figure \ref{fig:regions} allows the following types of orbits:

\begin{itemize}
\item[$\bullet$]{\it{Escape orbits}}. A particle moving in the yellow or cyan region, would approach at a minimum distance from the black hole and eventually would escape toward the cosmological horizon. On the other hand, a test particle moving in the gray region, depending on the initial conditions, will be either captured by the black hole or it will escape towards the cosmological horizon. 

\item[$\bullet$]{\it{Capture orbits}}. In this case the motion takes place in the blue region, the particle will be captured by the black hole. Also, an energy within the gray region may lead to a capture orbit. 

\item[$\bullet$]{\it{Bound orbits}}. A particle moving in the green region will follow a bound orbit, oscillating between two turning points, solutions of the equation $E=V_{eff}$. Some examples are represented in figures \ref{fig:orbitselectric}, \ref{fig:orbitselectric1} and \ref{fig:orbitselectric2}.

\item[$\bullet$]{\it{Circular orbits}}. The effective potential represented in the figure \ref{fig:regions} has two maxima (the black and red dots) and one minimum (the blue dot). Thus, two unstable circular orbits and a stable one may exist. 

\end{itemize}

In Figures \ref{fig:orbitselectric}–\ref{fig:orbitselectric2}, bound orbits for different values of $w$ are shown, taking the shape of hypotrochoids. In the figures 5.c, 6.c and 7.c, the orbits are quasi-circular. The deformation of the orbits arises from the electric interaction between the charged particle and the electrically charged black hole. The same behavior is observed in the  Reissner-Nordstrom de Sitter spacetime \cite{Olivares:2011xb}. 

\subsection{Circular orbits}

As noticed in the previous section, depending on the shape of the effective potential, one may have stable or unstable circular orbits. These can be found by imposing the conditions: $E=V_{eff}$ and $\frac{dV_{eff}}{dr}=0$. The second condition  leads to the equation of specific angular momentum:
\begin{equation}
4\Sigma^2L^4-r^3\Lambda^2\left(\varepsilon^2U^2rff'{^2}-4\Sigma \sigma\right)L^2-r^6\Lambda^4\left(\varepsilon^2U^2ff'{^2}-\sigma^2\right)=0,
\label{Leq}
\end{equation}
where we have used the notations:
\begin{equation}
\Sigma=rf\Lambda'-\frac{\Lambda}{4}(rf'-2f), \qquad \sigma=f\Lambda'-\frac{1}{2}f'\Lambda
\end{equation}
and $^\prime$ means the derivative with respect to $r$.
The discriminant of the equation being:
\begin{equation}
\Delta=\varepsilon^2U^2r^6\Lambda^4f f'{^2}\left(\varepsilon^2U^2r^2ff'{^2}+16\Sigma^2-8\Sigma\sigma r\right),
\end{equation}
one may write the solutions of (\ref{Leq}) in the form:
\begin{equation}
L_{\pm}^2=\frac{r^3\Lambda^2}{8\Sigma^2}\left[\varepsilon^2U^2rff'{^2}\left(1\pm \sqrt{1+\frac{8\Sigma (2\Sigma-\sigma r)}{f(\varepsilon Urf')^2}}\right)-4\Sigma \sigma\right].
\label{Lsol}
\end{equation}
In order to have physical orbits, one has to impose $\Delta \geq 0$ and the existence of at least one positive root defined in (\ref{Lsol}).
By introducing the notations:
\begin{eqnarray}
&& f_1 =\frac{2f\Lambda}{r(3U^2f+1)}, \quad
f_2 =\frac{4f\Lambda}{r(5U^2f+1)}, \quad
f_3 =\frac{\Lambda}{U^2r}
\end{eqnarray}
and
\begin{equation}
\varepsilon_1=\frac{U^2f+1}{2U\sqrt{f}}, \quad \varepsilon_2=\frac{1}{Uf'r}\sqrt{2(f'r-2f)+2U^2(4f^2+4ff'r-f'{^2}r^2)-2U^4f(2f^2+5ff'r+3f'{^2}r^2)},
\end{equation}
one obtains the cases presented in table \ref{tab:table2}. 

\begin{table}[h!]
\centering
\renewcommand{\arraystretch}{1.3}
\begin{tabularx}{\textwidth}{|c|Y|Y|}
\hline
{\it{$f'$}} & \textbf{$L^2$ } & \textbf{$\varepsilon$ }\\
\hline

$f'<0$ & 
$\begin{aligned}
L_+^2>0
\end{aligned}$
& 
$\begin{aligned}
|\varepsilon|>\varepsilon_1
\end{aligned}$\\
\cline{1-3}

\multirow{2}{*}{$0<f'<f_1$} 
& 
$\begin{aligned}
L_+^2>0
\end{aligned}$ 
& 
$\begin{aligned}
\forall \varepsilon
\end{aligned}$\\
\cline{2-3}

& 
$\begin{aligned}
L_-^2>0
\end{aligned}$ 
& 
$\begin{aligned}
|\varepsilon|<\varepsilon_1
\end{aligned}$\\
\cline{1-3}

\multirow{2}{*}{$f_1<f'<f_2$} 
& 
$\begin{aligned}
L_+^2>0
\end{aligned}$ 
& 
$\begin{aligned}
|\varepsilon|>\varepsilon_2
\end{aligned}$\\
\cline{2-3}

& 
$\begin{aligned}
L_-^2>0
\end{aligned}$ 
& 
$\begin{aligned}
\varepsilon_2<|\varepsilon|<\varepsilon_1
\end{aligned}$\\
\cline{1-3}

$f_2<f'<f_3$ & 
$\begin{aligned}
L_+^2>0
\end{aligned}$
& 
$\begin{aligned}
|\varepsilon|>\varepsilon_1
\end{aligned}$\\
\cline{1-3}

\end{tabularx}
\caption{Conditions on $f'$ and $\varepsilon$ for which $L_+^2$ and $L_-^2$ are positive quantities. }
\label{tab:table2}
\end{table}

For example, one may notice that particles with $|\varepsilon|>\varepsilon_1$ have only $L_+^2>0$, while particles with $|\varepsilon|<\varepsilon_1$ have both $L_+>0$ and $L_-^2>0$ for $0<f'<f_1$.

 At the radial coordinate $r_0=\left(-\frac{k(1+3w)}{2M}\right)^{1/3w}$, which is the solution of the equation $f'=0$, the angular momenta $L_+$ and $L_-$ vanish and the trajectory is radial. Thus, the radial coordinate $r_0$ separates the region where only $L_+^2>0$ from the one where both $L_+^2$ and $L_-^2$ can be positive quantities. 

\begin{figure}[H]
    \centering
    \begin{subfigure}{0.49\textwidth}
        \centering
        \includegraphics[scale=0.5, trim=2cm 10cm 0cm 1cm, clip]{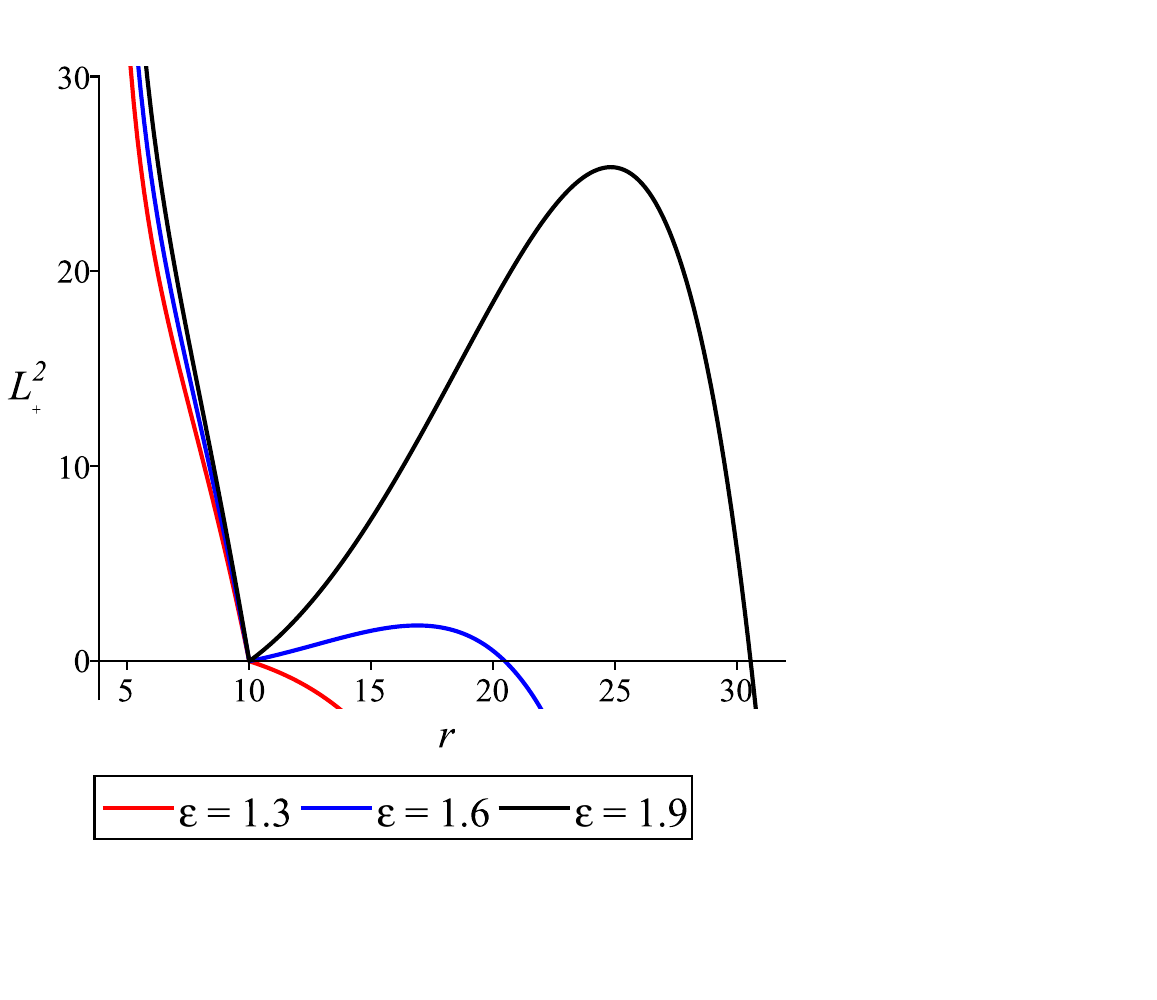}
    \end{subfigure}
    \hfill
    \begin{subfigure}{0.49\textwidth}
        \centering
        \includegraphics[scale=0.5, trim=0cm 10cm 0cm 1cm, clip]{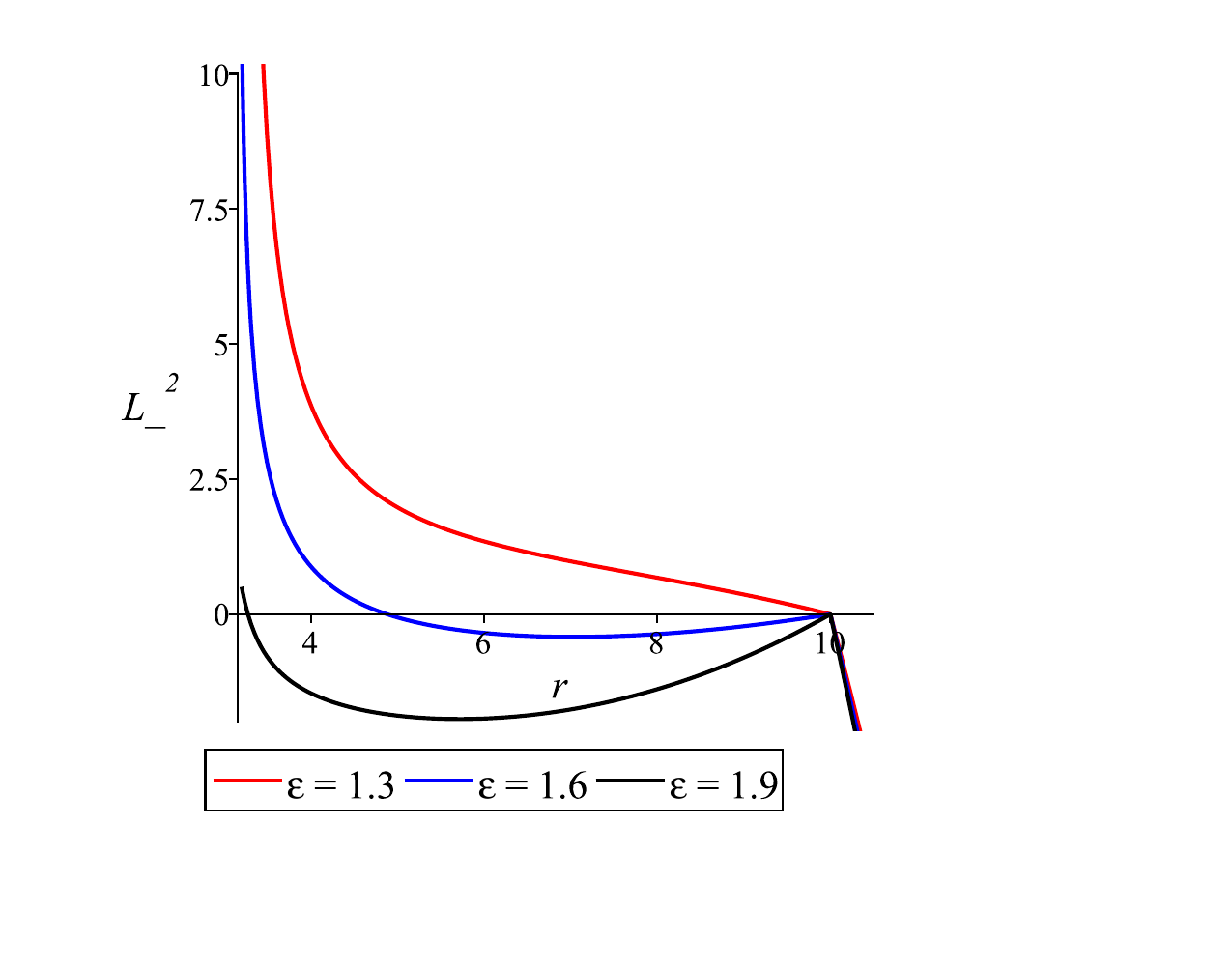}
    \end{subfigure}
    \caption{Plot of $L_+^2$ ({\it{left panel}}) and $L_-^2$ ({\it{right panel}}) for different values of $\varepsilon$. The parameters are: $M=1$, $U=0.5$, $k=0.02$ and $w=-2/3$.}
    \label{fig:LU}
\end{figure}

However, the constrains in the table \ref{tab:table2} are obtained for the general expression of $f$ given in (\ref{metric}) and $0<U<1$. Once we fix the values of the parameters $w$ and $k$, one can draw plots for $L_{\pm}^2$ as the ones given in the figure \ref{fig:LU}. As it can be noticed, for $r>r_0 = \sqrt{\frac{2M}{k}}$, which corresponds to the condition $f'<0$, the root $L_-^2$ is always negative while $L_+^2$ is positive for $|\varepsilon|>\varepsilon_1$.

The particle energy corresponding to a circular orbit is obtained by substituting the angular momenta (\ref{Lsol}) into the effective potential (\ref{potential}) and one obtains:
\begin{equation}
E_{\pm}=\frac{\varepsilon Uf}{\Lambda}+\sqrt{\frac{f}{\Lambda^2}\left(1+\frac{L_{\pm}^2}{\Lambda^2 r^2}\right)},
\label{Esol}
\end{equation}
which should be positive quantities. In the case of $\varepsilon>0$, the energies $E_{\pm}$ are always positive. However, for $\varepsilon<0$, one has to impose $|\varepsilon|<\frac{1}{U\sqrt{f}}$. 

A special case occurs when the angular momenta coincide, i.e. $L_-^2=L_+^2$. By imposing the condition  $\Delta=0$, one obtains the following equation for the corresponding radius $r_0$:
\begin{align*}
\varepsilon^2 U^2 ff'{^2}r^2+16\Sigma^2-8\Sigma \sigma r=0.
\end{align*}

\begin{figure}[H]
    \centering
    \begin{subfigure}{0.49\textwidth}
        \centering
        \includegraphics[scale=0.5, trim=2cm 10cm 0cm 1cm, clip]{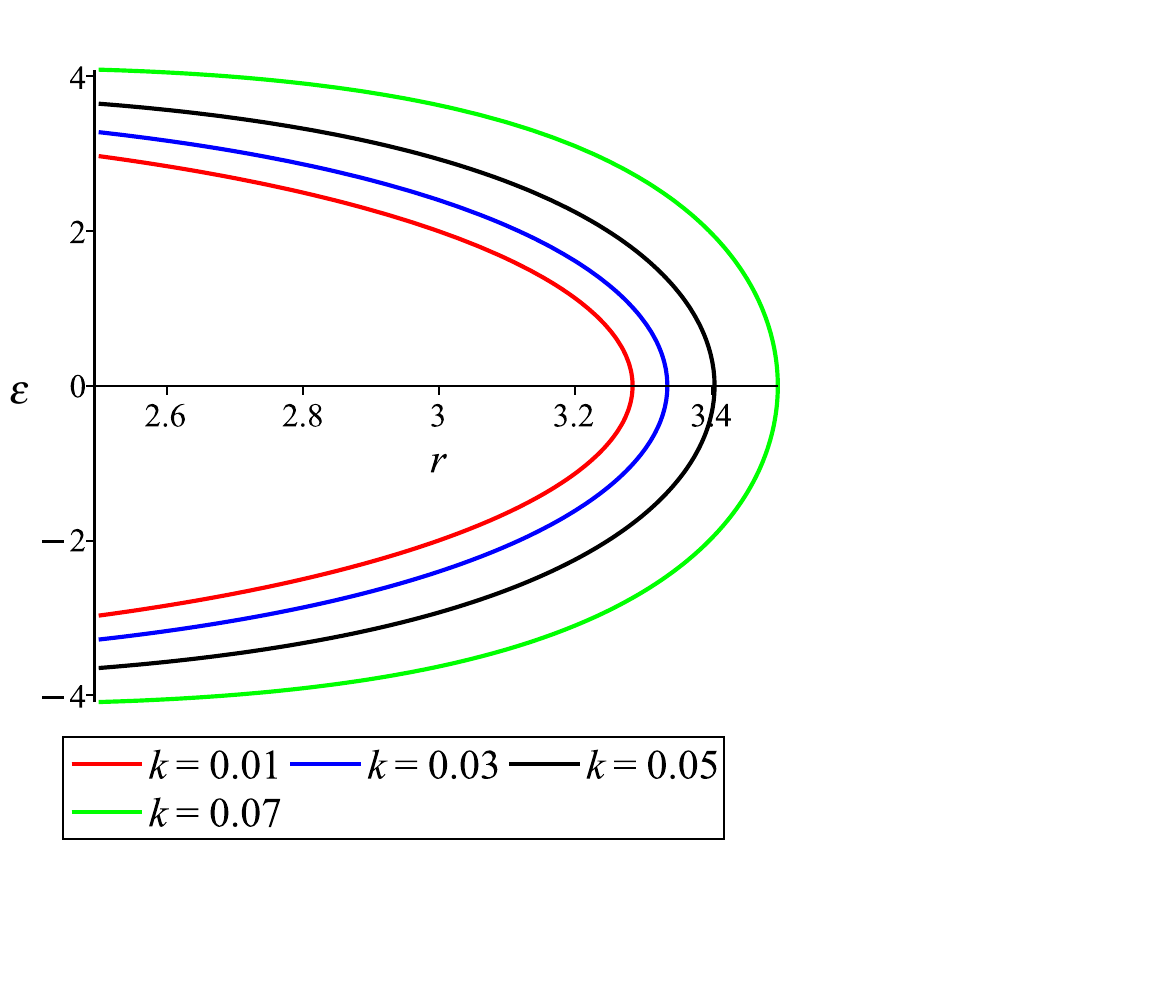}
    \end{subfigure}
    \caption{Implicit plot of curves for which $L_-^2=L_+^2$. The value of parameters are: $M=1$, $w=-2/3$ and $U=0.4$.}
    \label{fig:Lequal}
\end{figure}

As it can be noticed in the figure \ref{fig:Lequal}, the radius for which the values of angular momenta coincide increases with $k$. On the other hand, for a given value of $k$, the radius decreases with $|\varepsilon|$
and is reaching the maximum value for $\varepsilon=0$.

\begin{figure}[H]
    \centering
    \begin{subfigure}{0.49\textwidth}
        \centering
        \includegraphics[scale=0.5, trim=2cm 10cm 0cm 1cm, clip]{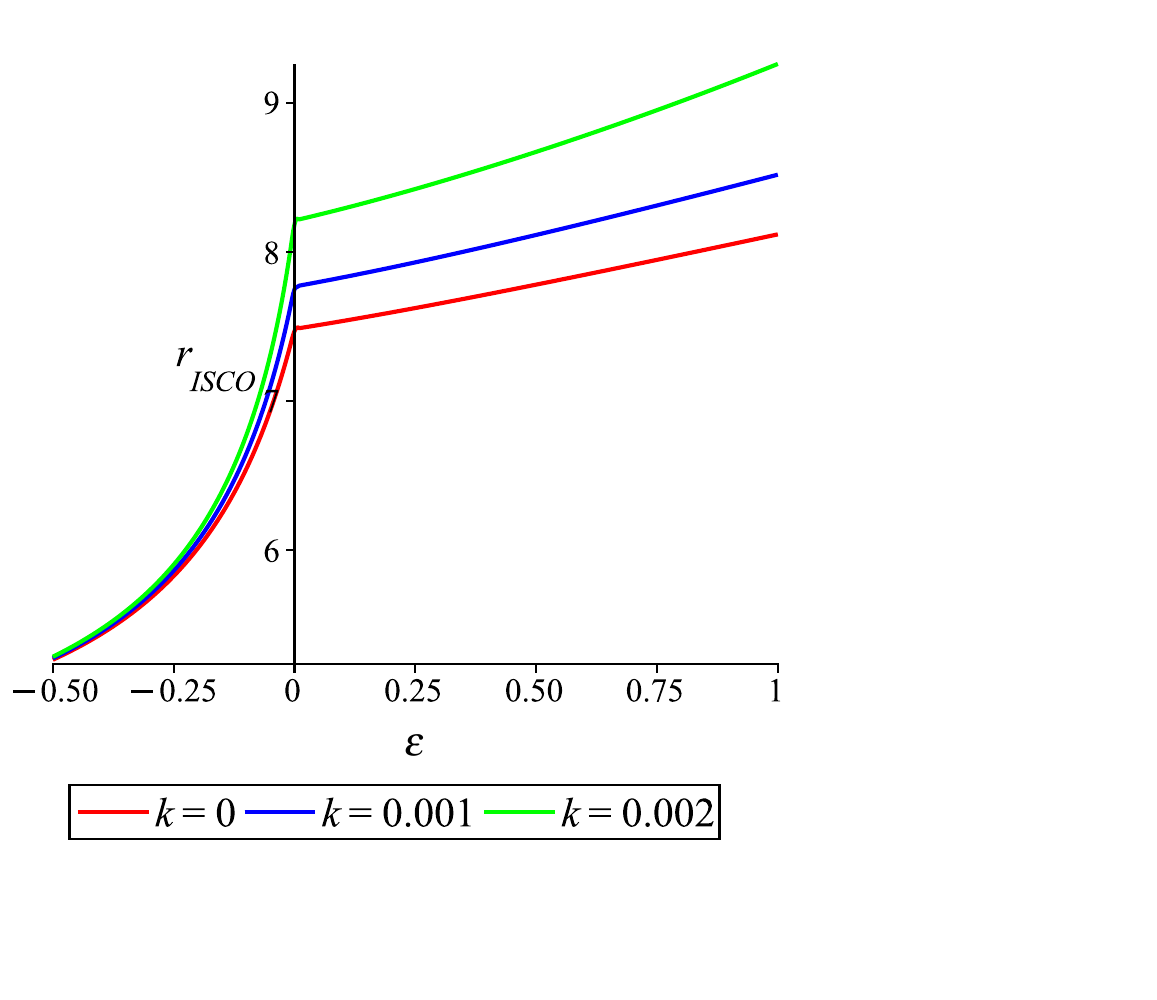}
    \end{subfigure}
    \caption{Dependence of $r_{ISCO}$ in terms of $\varepsilon$ for different values of $k$. The values of the parameters are: $M=1$, $w=-2/3$ and $U=0.5$.}
    \label{fig:risco}
\end{figure}

The innermost stable circular orbit (ISCO) is the smallest marginally stable circular orbit followed by a test particle which is orbiting the black hole. Its location is given by the inflection point of the effective potential, $V_{eff}''=0$. By solving the equations $V_{eff}'=0$ and $V_{eff}''=0$, one finds that the specific charge and angular momentum for this orbit satisfy the following relations:
\begin{equation}
\varepsilon=\frac{\Lambda^2r^3\left(2\Lambda'f-\Lambda f'\right)-\left(\Lambda f'r-4\Lambda'fr-2\Lambda f\right)L_{ISCO}^2}{2U \Lambda r^2 \left(\Lambda f'-\Lambda' f\right) \sqrt{f\left(\Lambda^2 r^2+L_{ISCO}^2\right)}},
\label{epsilonisco}
\end{equation}
\begin{equation}
a L_{ISCO}^4+b L_{ISCO}^2-f'{^2}=0,
\label{Lisco}
\end{equation}
where we introduced the notations:
\begin{align*}
a=\frac{1}{f'\Lambda^{6} r^6 }\lbrace 4f^2f''\Lambda^2 r+r^2[3U^2f(U^2f+2)-1]f'{^3}-4f\Lambda r(U^2f+1)f'{^2}+8f^2f'\Lambda^2 \rbrace
\end{align*}
\begin{equation}
b=\frac{2}{f'\Lambda^{4} r^4 }\lbrace 2f^2f''\Lambda^2 r+r^2[U^2f(3U^2f+4)-1]f'{^3}-2f\Lambda r(3U^2f+1)f'{^2}+6f^2f'\Lambda^2 \rbrace
\end{equation}
Once one obtains the solution of the equation (\ref{Lisco}), this can be substituted back into the equation (\ref{epsilonisco}) in order to find the radius of the innermost stable circular orbit, $r_{ISCO}$. For the general metric (\ref{metric}), this can be obtained only numerically. For specific values of the model's  parameters, $r_{ISCO}$ is represented in the figure \ref{fig:risco}. As it can be noticed, for negative values of the specific charge $\varepsilon$, the radius of ISCO is decreasing as $|\varepsilon|$ increases as a result of the attractive Coulomb interaction with the black hole's charge. On the other hand, for positively charged particles, the ISCO radius shifts further away from the black hole. In both cases, the presence of quintessence leads to an increase of $r_{ISCO}$ pointing out a repulsive force induced by the quintessential fluid.  

\subsection{The Lyapunov exponent and stability of circular orbits}

Let us consider now the case of a particle with unit mass whose motion is confined to the equatorial plane. The corresponding Hamiltonian derived from (\ref{H1}) leads to the equations of motion:
\begin{eqnarray}
&&
\dot{r} = \frac{\partial H}{\partial \pi_r} = \frac{f}{\Lambda^2} \pi_r,~~~\dot{t} = \frac{\partial H}{\partial \pi_t} = \frac{\Lambda^2}{f} (E+ \varepsilon A_t),\nonumber \\*
&&
\dot{\pi}_r = - \frac{\partial H}{\partial r}  = - \frac{1}{2}
\left[ \left( \frac{f}{\Lambda^2} \right)^{\prime} \pi_r^2 - \frac{(\Lambda^2 r^2)^ {\prime}}{\Lambda^4 r^4} L^2 - \left( \frac{\Lambda^2}{f} \right)^{\prime} (E+ \varepsilon A_t)^2
-2 \varepsilon \frac{\Lambda^2}{f} (E+ \varepsilon A_t) A_t^{\prime}  \right],
\end{eqnarray}
where we have introduced the conserved quantities $E$ and $L$.
Now, one may use the general procedure described in \cite{Gao:2022ybw} in detail to write down the following relations:
\begin{eqnarray}
&&
\frac{dr}{dt} = \frac{\dot{r}}{\dot{t}} = \frac{f^2}{\Lambda^4} \frac{\pi_r}{E+ \varepsilon A_t} \equiv F (r , \pi_r ),
\nonumber \\*
&&
\frac{d\pi_r}{dt} = \frac{\dot{\pi}_r}{\dot{t}} = \varepsilon A_t^{\prime} - \frac{f}{2 \Lambda^2 (E+ \varepsilon A_t)} 
\left[ 
 \left( \frac{f}{\Lambda^2} \right)^{\prime} \pi_r^2 - \frac{(\Lambda^2 r^2)^ {\prime}}{\Lambda^4 r^4} L^2 \right]
+ \frac{f}{2 \Lambda^2} \left( \frac{\Lambda^2}{f} \right)^{\prime} (E+ \varepsilon A_t)  \equiv G (r , \pi_r )
\nonumber \\*
\label{FG}
\end{eqnarray}
where $E+ \varepsilon A_t$ should be replaced with the constraint (\ref{firstintegral}) written as:
\[
E + \varepsilon A_t = \sqrt{ \frac{f}{\Lambda^2} + \frac{f^2}{\Lambda^4} p_r^2 + \frac{f}{r^2 \Lambda^4} L^2 }.
\]
The two functions $F$ and $G$ introduced in (\ref{FG}) will be used to construct in the phase space $(r , \pi_r)$ the Jacobian matrix whose elements are defined as:
\[
K_{11} = \frac{\partial F}{\partial r}
\; , \quad
K_{12} = \frac{\partial F}{\partial \pi_r}
\; , \quad
K_{21} = \frac{\partial G}{\partial r}
\; , \quad K_{22} = \frac{\partial G}{\partial \pi_r}.
\]
Since the motion is circular, one has $\pi_r =0$ and therefore
the elements of the matrix $K_{ij}$ have the explicit expressions :
\begin{eqnarray}
&&
K_{11} = 0 \; , \quad K_{12} = \left( \frac{f}{\Lambda^2} \right)^{3/2}  \left[ 1 + \frac{L^2}{r^2 \Lambda^2} \right]^{-1/2}
\nonumber \\*
&&
K_{21} = \varepsilon A_t^{\prime \prime}
-  \frac{1}{2} \left[ \left( \frac{f}{\Lambda^2} \right)^{\prime \prime} + L^2 \left( \frac{f}{r^2 \Lambda^4} \right)^{\prime \prime} \right]
\left[  \frac{f}{\Lambda^2} \left( 1 + \frac{L^2}{r^2 \Lambda^2} \right) \right]^{-1/2}
\nonumber \\*
&&
+ \frac{1}{4} \left[ \left( \frac{f}{\Lambda^2} \right)^{\prime} + L^2 \left( \frac{f}{r^2 \Lambda^4} \right)^{\prime} \right]^2
\left[  \frac{f}{\Lambda^2} \left( 1 + \frac{L^2}{r^2 \Lambda^2} \right) \right]^{-3/2} \; ,
\quad K_{22} = 0.
\end{eqnarray}
These allow us to define the Lyapunov exponent as being given by:
$\lambda^2 = K_{12} K_{21}$, i.e.
\begin{eqnarray}
\lambda^2 =
&&
- \varepsilon U \left( \frac{f}{\Lambda} \right)^{\prime \prime} \frac{f^{3/2}}{\Lambda^3}  \left[ 1 + \frac{L^2}{r^2 \Lambda^2} \right]^{-1/2}
- \frac{f}{2 \Lambda^2} \left[ \left( \frac{f}{\Lambda^2} \right)^{\prime \prime} + L^2 \left( \frac{f}{r^2 \Lambda^4} \right)^{\prime \prime} \right]
\left[ 1 + \frac{L^2}{r^2 \Lambda^2} \right]^{-1}+
\nonumber \\*
&&
+ \, \frac{1}{4} \left[ \left( \frac{f}{\Lambda^2} \right)^{\prime} + L^2 \left( \frac{f}{r^2 \Lambda^4} \right)^{\prime} \right]^2
\left[ 1 + \frac{L^2}{r^2 \Lambda^2 } \right]^{-2} .
\label{lyapunov}
\end{eqnarray}

One may notice that $\lambda^2$ is depending on the parameters $M$, $k$, $w$ and $U$ and its numerical value can be obtained only numerically. Once the model's parameters together with the particle's charge and angular momentum are fixed, the radius of the circular orbit is a numerical solution of the equation (\ref{Leq}). This value will be replaced in the relation (\ref{lyapunov}) for deriving the numerical value of the Lyapunov exponent which characterizes the stability of the timelike circular orbit. However, one has to choose from the large number of roots of the equation (\ref{Leq}) the solution corresponding to the maximum or minimum value of the effective potential.
If $\lambda^2$ is positive, the circular motion is unstable \cite{Gao:2022ybw}. 

\subsection{Unstable circular orbits}

\begin{figure}[H]
\centering
\includegraphics[width=0.45\textwidth]{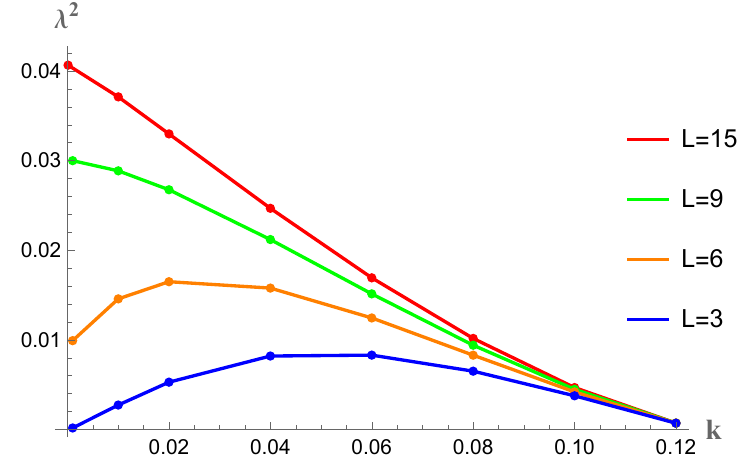}
\includegraphics[width=0.45\textwidth]{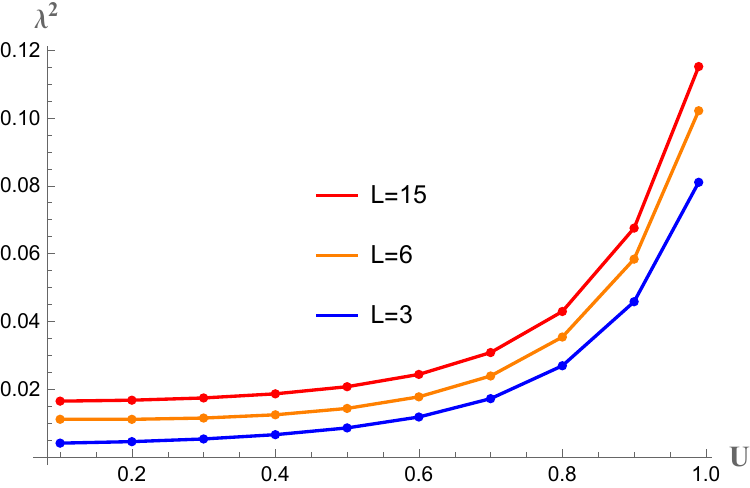}
\caption{{\it Left panel}. The Lyapunov exponent (\ref{lyapunov}) as a function of $k$, for  $U=0.5$ and different values of the angular momentum. {\it Right panel}. The Lyapunov exponent (\ref{lyapunov}) as a function of $U$, for  $k=0.05$ and different values of the angular momentum.The other numerical values are: $M=1$, $w=-2/3$ and $\varepsilon =1$.}
\label{Ly}
\end{figure}

As it has been seen in the section 3 devoted to the analysis of the effective potential, since the potential vanishes on the horizons and has at least one maximum value inbetween, there always exists an unstable circular orbit situated just outside the event horizon. In the followings, let us compute the Lyapunov exponent for such an orbit using a numerical approach.
As an illustrative example, we consider the particular value $w=-2/3$ and represent in the figure \ref{Ly}, the Lyapunov exponent $\lambda^2$ defined in (\ref{lyapunov}), for different values of the angular momentum, as a function of $k$ in the left panel and as a function of $U$ in the right panel. Since we focus on an unstable orbit, the values of $\lambda^2$ are always positive.
Once the parameters $k$ and $U$ are fixed, the angular momentum of the charged particle affects the effective potential and the position of the circular orbit given by the equation (\ref{Leq}). As expected, the value of the Lyapunov exponent increases with $L$ and this behavior is more prominent for small values of $k$. By inspecting the plots in the right panel, one can see that, for given values of $k$ and $L$, the Lyapunov exponent is monotonically increasing with $U$ and tends to its maximum value when $U \to 1$. Let us also notice in the left panel that the shape of $\lambda^2$ is depending on the value of the angular momentum. Thus, for $L=3$, the Lyapunov exponent has a maximum value for $k$ close to the middle of its allowed range. As $L$ increases, the maximum of $\lambda^2$ is strongly increasing and moves to smaller values of $k$. There is a critical value of the angular momentum above which the Lyapunov exponent is monotonically decreasing with $k$ (see the red and green plots in the left panel).

\subsection{Stable circular orbits}

As it has been noticed in the figures \ref{fig:potU} and \ref{fig:potk}, for specific values of the parameters $w$, $k$ and $U$, a second maximum of the potential may appear close to the cosmological horizon. The charged particle with suitable $\varepsilon$ and $L$ moves on a stable circular orbit whose radius is one of the solutions of the equation (\ref{Leq}). 

\begin{figure}[H]
\centering
\includegraphics[width=0.45\textwidth]{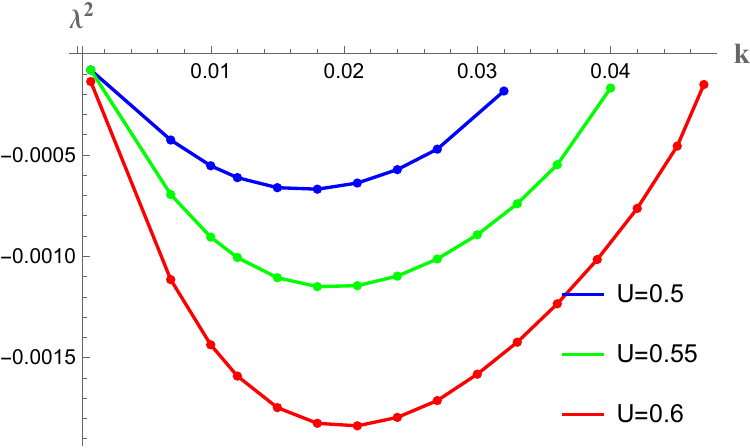}
\caption{The Lyapunov exponent (\ref{lyapunov}) as a function of $k$, for different values of $U$. The other numerical values are: $M=1$, $w=-2/3$, $L=3$ and $\varepsilon =-2$.}
\label{LyNeg}
\end{figure}

In the figure \ref{LyNeg}, one may notice that the Lyapunov exponent and the range of $k$ that allows stable circular orbits are very sensitive to the values of the parameter $U$. Thus, for $U=0.5$, the stable orbits are allowed for very small values of $k$ and once $k$ increases, the circular orbits become unstable. As expected for particles with negative $\varepsilon$, when the charge of the black hole increases, the circular orbits become more stable.

\subsection{Charged particles with zero angular momentum}

In the particular case of test particles with $L=0$, the effective potential (\ref{potential}) reduces to the simple expression:
\begin{equation}
V_0=\frac{\sqrt{f}}{\Lambda}\left(\varepsilon U \sqrt{f}+1\right).
\label{V0}
\end{equation}
An interesting behavior is that the charged test particles may have bound orbits while moving radially. This is due to the repulsive term $\varepsilon U$, with $\varepsilon<0$, which can balance the gravity even without an angular momentum. In the Kiselev case this behavior is not possible, since the quintessence term cannot compensate the attractive gravitational contribution.

\begin{figure}[H]
    \centering
    \begin{subfigure}{0.49\textwidth}
        \centering
        \includegraphics[scale=0.5, trim=2cm 10cm 5cm 1cm, clip]{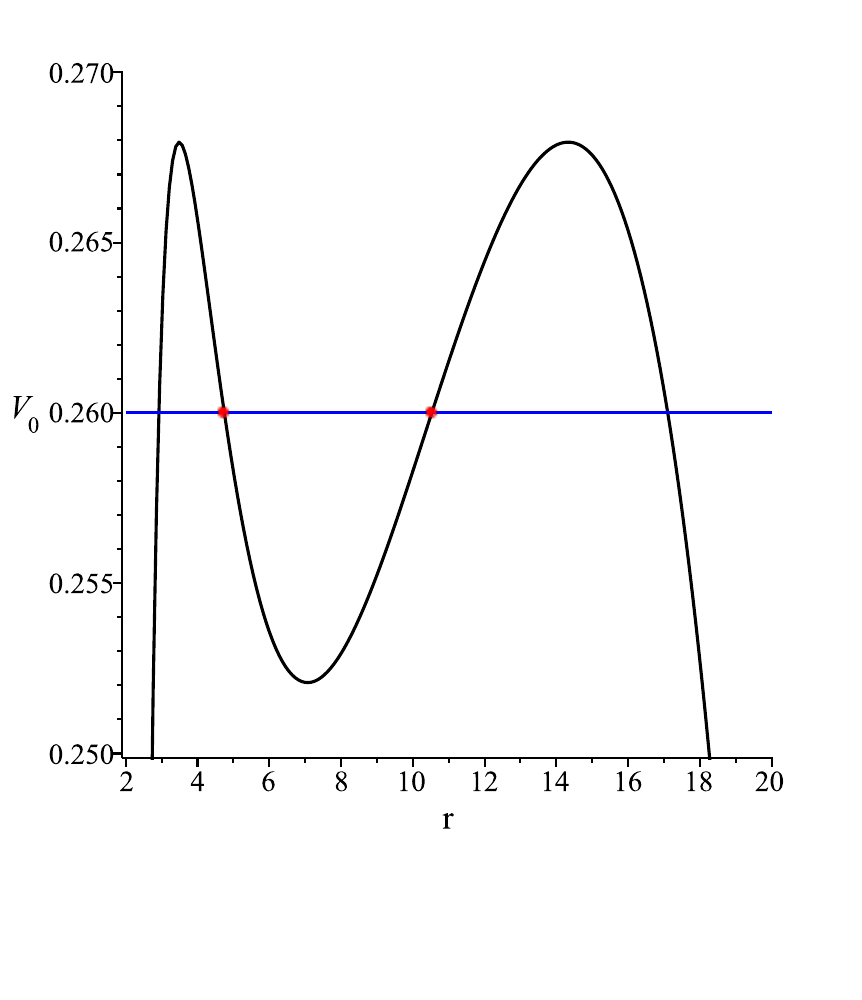}
    \end{subfigure}
    \hfill
    \begin{subfigure}{0.49\textwidth}
        \centering
        \includegraphics[scale=0.6, trim=2cm 12cm 0cm 1cm, clip]{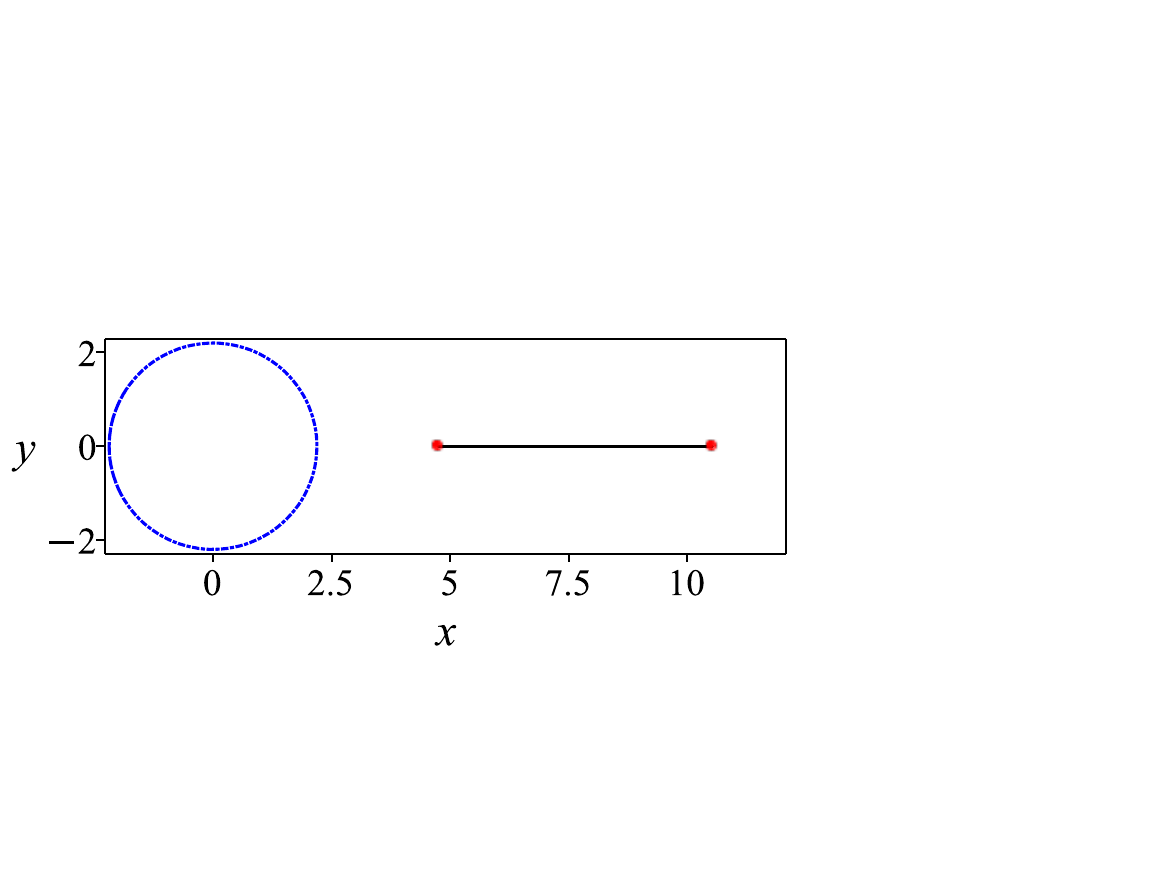}
    \end{subfigure}
    \caption{{\it{Left panel}}. The effective potential (\ref{V0}).The particle's energy is represented by the blue horizontal line. {\it{Right panel}}.  The black horizontal line represents a bound radial geodesic of a charged particle with $L=0$ oscillating between two turning points represented by the red dots. The dashed blue circle corresponds to the black hole's horizon. The values of the other parameters are: $M=1$, $w=-2/3$, $U=0.5$, $k=0.04$, $\varepsilon=-2$ and $E=0.26$}. 
    \label{fig:radialtraj}
\end{figure}

For the particle moving on a {\it{circular orbit}} with $L=0$, one has to impose the condition $V_0'=0$, which leads to the relation
\begin{equation}
f^{\prime} \left[ 1+ 2 \varepsilon U \sqrt{f} + U^2 f \right] =0.
\label{fp}
\end{equation}
The solution of the equation $1+ 2 \varepsilon U \sqrt{f} + U^2 f =0$ corresponds to the negative specific charge: 
\begin{equation}
\varepsilon_0=\frac{-(1+U^2f)}{2U\sqrt{f}}.
\label{eo}
\end{equation}
By substituting this expression into the second derivative of the potential, one obtains:
\begin{equation}
V_0''=-\frac{f'{^2}}{4f^{3/2}\Lambda},
\label{V0sec}
\end{equation}
which is always negative, pointing out an unstable orbit whose radius corresponds to the maximum value of the potential.

The other solution of the equation (\ref{fp}) corresponds to $f'=0$ and has the expression $r_0 =\left(-\frac{k(1+3w)}{2M}\right)^{1/3w}$, which is independent of the parameter $U$. Depending on the sign of
\[
V_0^{\prime \prime} = \left. \frac{f^{\prime \prime} [1+ 2 \varepsilon U \sqrt{f} + U^2 f ]}{2 \sqrt{f} \Lambda^2} \right|_{r=r_0},
\]
this orbit can be either stable or unstable. Since $f^{\prime \prime} = \frac{6wM}{r_0^3} <0$, it turns our that, unlike the situation for the Kiselev black hole, in our case, particles with $\varepsilon < \varepsilon_0$ with $\varepsilon_0$ given in (\ref{eo}) are able to follow stable orbits.

The term {\it circular orbit} used here is generic, since a particle with zero angular momentum cannot evolve on a circular trajectory around the black hole. Instead, for a stable orbit, the particle will stay at a specific position (the local minimum of the potential), corresponding to the radius of circular orbit \cite{Stuchlik:2002tj}. If the orbit is unstable, under a small perturbation, the particle will either fall in the black hole or escape to the cosmological horizon.

\section{The periapsis shift} 
 In this section, let us discuss the periapsis shift of a quasi-circular
orbit, which, as it was mentioned in the Introduction, has deep implications in correlating the theory with observational data.
For a slightly perturbed particle from its stable circular orbit, the shift occurs when the radial frequency $\omega_r$
is not equal to the orbital frequency $\omega_{\varphi}$ and it can be defined as \cite{Harada:2022uae}:
\begin{equation}
\Delta \phi_P = 2 \pi \left[ \frac{\omega_{\varphi}-\omega_r}{\omega_r} \right] = 2 \pi \left[ \frac{1}{\sqrt{A}} -1 \right],
\label{prec}
\end{equation}
where,
\begin{equation}
A = \left( \frac{\omega_r}{\omega_{\varphi}} \right)^2.
\label{Agen}
\end{equation}
In our case, the orbital frequency is given by the expression:
\begin{equation}
\omega_{\varphi} = \dot{\varphi} = \frac{L}{r^2 \Lambda^2},
\label{omegaphi}
\end{equation}
while the radial frequency can be obtained by employing the Hamiltonian formalism for epicyclic motion presented in \cite{Stuchlik:2021guq}.

Let us start with the case of an uncharged particle moving on a stable circular orbit $r_c$ whose potential (\ref{potential}) leads to the angular momentum and energy defined in (\ref{Lsol}) and (\ref{Esol}). The particle is oscillating around the circular orbit, $r = r_c + \delta r$, and the
small perturbation $\delta r$ is satisfying the equations of a linear harmonic
oscillation, $\delta \ddot{r} + \omega_r^2 \delta r = 0$,
where {\it dot} denotes the derivative with respect to the proper time. 

The radial frequency measured by a local observer as being \cite{Stuchlik:2021guq}:
\begin{equation}
\omega_r^2 = \frac{1}{g_{rr}} \frac{\partial^2 H_{pot}}{\partial r^2} = \frac{f}{\Lambda^2} \frac{\partial^2 H_{pot}}{\partial r^2} .
\label{omegar2}
\end{equation}
In the following we shall present a few simple cases and recover some results previously known in literature.

\subsection{The uncharged Kiselev black hole}

As an important physical example, let us discuss the particular case with $U=0$ for which the potential in (\ref{firstintegral}) becomes:
\begin{equation}
V = f\left(1+\frac{L^2}{r^2}\right)
\label{V}
\end{equation}
and the circular orbit condition $V^{\prime} (r_c) =0$ leads to the following expressions of the angular momentum
and energy:
\begin{equation}
L^2 
=  \frac{r^3  f^{\prime}  }{2f -r f^{\prime} } \; , \quad
E^2 = \frac{2 f^2  }{ 2f -r f^{\prime} }.
\label{L2E}
\end{equation}
Using the potential part of the Hamiltonian:
\[
H_{pot} = \frac{1}{2} \left[ - \frac{E^2}{f} + \frac{L^2}{r^2} \right]
\]
one computes the radial and angular frequencies as being given by:
\[
\omega_r^2 = f \frac{\partial^2 H_{pot}}{\partial r^2} = \frac{f}{2} \left[ \frac{E^2 (f f^{\prime \prime}-2f^{\prime2})}{f^3} + \frac{6L^2}{r^4} \right] , \quad
\omega_{\varphi} = \frac{L}{r^2}.
\]
Thus, one obtains the following expression of the quantity $A$ defined in (\ref{Agen}):
\begin{equation}
A = r f \left[ \frac{f^{\prime \prime}}{f^{\prime}} - 2 \frac{f^{\prime}}{f} + \frac{3}{r} \right], 
\label{A}
\end{equation}
which agrees with the expression derived in \cite{Harada:2022uae}, for a general spherically symmetric static metric.
Obviously, one has to impose the following conditions: 
\begin{equation}
f^{\prime} >0 \; , \quad 2f - r f^{\prime} >0 \; , \quad V^{\prime \prime} >0,
\label{cond}
\end{equation} 
so that $L^2$ and $E^2$ are positive and the circular orbit is stable. \\

If one replaces the function $f$ with the Kiselev expression (\ref{metric}), one finds the general expression of $A$ which is depending on the parameters $k$ and $w$ as:
\begin{equation}
A_w = \frac{-2M(6M-r)r^{6w} + (1-9w^2)k r^{3w+1} + 6kM(3w^2-4w-2)r^{3w} -3k^2(3w^2+4w+1)}{r^{3w+1} [2Mr^{3w}+(3w+1)k]}
\label{Aw}
\end{equation}
For the physical important value $w=-2/3$, for which the metric function is
(\ref{metric23}), the relation (\ref{Aw}) becomes:
\begin{equation}
A_{-\frac{2}{3}} = 
\frac{2Mr(1+6kr)-kr^3(3-kr)-12M^2}{r(2M-kr^2)}.
\label{Ak}
\end{equation}
Further, one finds that the conditions (\ref{cond}) impose a maximum value of the parameter $k$, i.e. $k_{max} = (3-2 \sqrt{2})/(32M)$. Moreover, the expression (\ref{Ak}) is always less than one, leading to a positive precession angle (\ref{prec}). \\

\subsection{The Reissner-Nordstr\"om and Schwarzschild cases}

In absence of the quintessence fluid, by using the coordinate transformation $R = \Lambda r$, the metric (\ref{metelec}) reduces to the Reissner-Nordstr\"om one \cite{Stelea:2023yqo}\footnote{Here, since the geometry is asymptotically flat one has to use $C=1-U^2$.}. In this case, the relation (\ref{A}) has the form:
\begin{equation}
A_{RN} = \frac{(r-6M)Mr^2 + 9MQ^2 r -4Q^4}{r^2(Mr-Q^2)}
\label{ARN}
\end{equation}
and it leads again to a prograde periapsis shift.
For $Q=0$, one recovers 
the well-known Schwarzschild case with:
\begin{equation}
A_S = \frac{r-6M}{r} \; , \quad
\Delta \phi_S = 2 \pi \left[ \frac{1}{\sqrt{1-\frac{6M}{r}}} -1 \right] \approx \frac{6 \pi M}{r}
\label{As}
\end{equation}
For $r>6M$ and $Q^2 \ll M^2$, one has:
\[
\Delta \phi_{RN} \approx  \frac{6 \pi M}{r} - \frac{\pi Q^2}{Mr} < \Delta \phi_S.
\]
The above result agrees with the one derived in \cite{Avalos-Vargas:2012jja}, for the orbital motion for charged particles moving in
the equatorial plane of the Reissner-Nordstr\"om source. Thus, the charge $Q$ does decrease the periapsis advance, which remains however still positive.

On the other hand, to first order in $M/r$ and $kr$ i.e. $3M \ll r \ll 1/k$, the relation (\ref{Ak}) becomes:
\[
A_{-\frac{2}{3}} \approx \frac{r-6M}{r} - \frac{kr(r-3M)}{M}  < A_S,
\]
leading to:
\[
\Delta \phi_{-\frac{2}{3}} \approx \frac{6 \pi M}{r} + \frac{\pi kr ( r-3M)}{M} > \Delta \phi_S.
\]
For $r>3M$, one may notice that the presence of the parameter $k$ is increasing the value of $\Delta \phi$,
compared to the Schwarzschild or Reissner--Nordstr\"om cases. Thus, one may expect that, for a Kiselev black hole, the periapsis advance is always positive. 

\subsection{Uncharged particles around an electric Kiselev black hole}

Let us turn now to the slightly more complicated case of an uncharged particle evolving around an electrically charged Kiselev black hole. The Hamiltonian
(\ref{HpotEl}) written for $\varepsilon =0$, taking into account the relations
(\ref{L2E}), allows us to compute the radial and azimuthal frequencies defined in (\ref{omegar2})  and (\ref{omegaphi}).
Putting everything together, we obtain the expression of $A$ defined in (\ref{Agen}) as being:
\begin{eqnarray}
A_U & = & \frac{1}{\Lambda_e^2(1+xf)f^{\prime}} \left \lbrace x^3 f^2 \left[ 
r f^2 f^{\prime \prime} + 3r^2 f^{\prime 3} + 3 f^2 f^{\prime} + 4r f f^{\prime 2} \right] 
- x^2 f \left[ 
r f^2 f^{\prime \prime} - 3r^2 f^{\prime 3} + 3 f^2 f^{\prime} \right]  \right.
\nonumber \\*
 & & \left. -x \left[ 
r f^2 f^{\prime \prime} - 2r^2 f^{\prime 3} + 2r f f^{\prime 2} + 3 f^2 f^{\prime} \right] + r f f^{\prime \prime} -2r f^{\prime 2} + 3 f f^{\prime} \right \rbrace
\end{eqnarray}
where $x = U^2$.
To first order in $x$ and $k$, the above expression, with the metric function (\ref{metric23}), becomes
\[
A_U \approx \frac{r-6M}{r} - \left( \frac{kr}{M} + \frac{8Mx}{r^2} \right) ( r-3M) = A_k - \frac{8Mx(r-3M)}{r^2},
\]
leading to an increased periapsis shift of the charged Kiselev solution compared to the uncharged one.

\begin{figure}[H]
  \centering
  \includegraphics[width=0.45\textwidth]{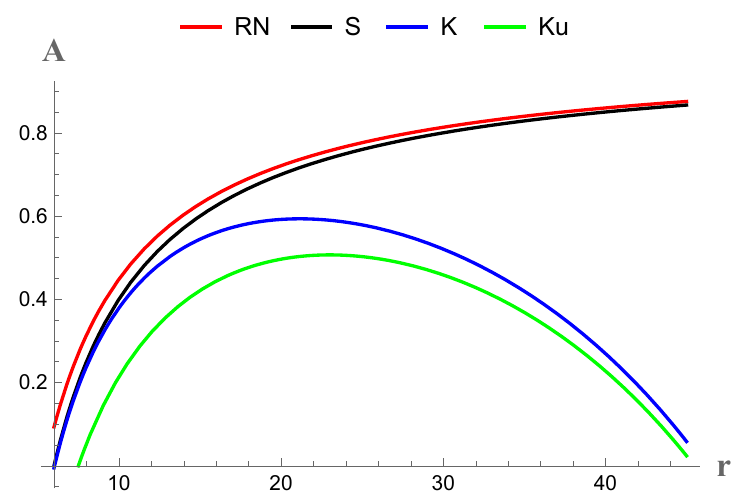}  
\includegraphics[width=0.45\textwidth]{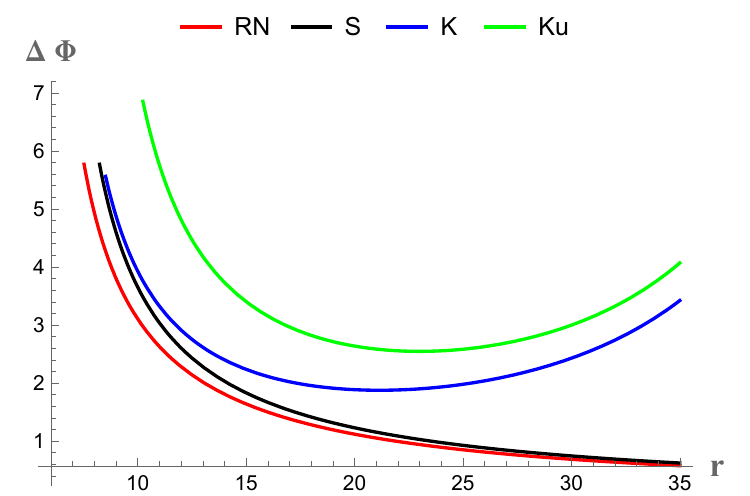}
  \caption{The expressions of $A$ and $\Delta \phi$ for: Schwarzschild (S), Reissner-Nordstrom (RN), Kiselev (K) and chargeless particle in charged Kiselev (Ku). The numerical values are: $M=1$, $Q=0.6$, $w=-2/3$, $x=0.3$, $k=0.0003$
.} 
  \label{AFig}
\end{figure}

In order to summarize the results presented above, in Figure \ref{AFig}, we are representing the expressions of $A$ and the expressions of $\Delta \phi$, as functions of $r$, for the cases discussed above, namely: Schwarzschild (S), Reissner-Nordstr\"om (RN), Kiselev (K) and uncharged particles in electrically charged Kiselev (Ku).
One may notice that, for large values of $r$, the behavior of $A$ and $\Delta \phi$ deviate significantly from the Schwarzschild and Reissner-Nordstr\"om cases where $A \to 1$ and $\Delta \phi \to 0$, for $r \to \infty$.
For a fixed value of $r$ in the allowed range for bounded trajectories, the presence of quintessence is increasing the value of $\Delta \phi$. 
Also, unlike the Schwarzschild and Reissner-Nordstr\"om cases, there is a value of $r$, situated between the horizons, where $\Delta \phi$ has a minimum value and is increasing afterwards. 

\subsubsection {Charged particles around the electrically charged Kiselev black hole}

The case of a charged particle is much more complicated and one has to start with the potential part of the Hamiltonian as defined in (\ref{HpotEl}) and compute the radial frequency defined in (\ref{omegar2}) as:
\begin{eqnarray}
\omega_r^2 & = & \frac{f}{2 \Lambda^2} \left \lbrace \frac{L^2}{r^4 \Lambda^4} \left[
6 \Lambda^2+8r \Lambda \Lambda^{\prime} + 6 r^2 \Lambda^{\prime 2} - 2 r^2 \Lambda \Lambda^{\prime \prime} \right]
\right \rbrace \nonumber \\*
& + &
\left. \frac{E^2}{f^3} \left[ 4 f f^{\prime} \Lambda \Lambda^{\prime} - 2 f^{\prime 2} \Lambda^2 - 2 f^2 \Lambda^{\prime 2} + f f^{\prime \prime} \Lambda^2 -2 f^2 \Lambda \Lambda^{\prime \prime} \right] + 2 \varepsilon U E \Lambda^{\prime \prime} - \varepsilon^2 U^2 f^{\prime \prime} \right \rbrace,
\end{eqnarray}
while the angular frequency is $\omega_{\varphi} = L/(r^2 \Lambda^2 )$.
In these expressions, the angular momentum $L$ is deined in (\ref{Lsol}), while the energy of the particle moving on the circular orbit is defined in ({\ref{Esol}). The expression of the periapsis shift can be numerically evaluated using (\ref{prec}). Depending on the parameters values, it can be either positive (for $A<1$) or negative (for $A>1$), as it has been noticed in Figure \ref{A12}.

This indicates the possibility of prograde or retrograde precessions for charged particles in the electrically charged Kiselev black hole.

\begin{figure}[H]
  \centering
  \includegraphics[scale=0.4,trim = 1cm 9cm 2cm 1cm]{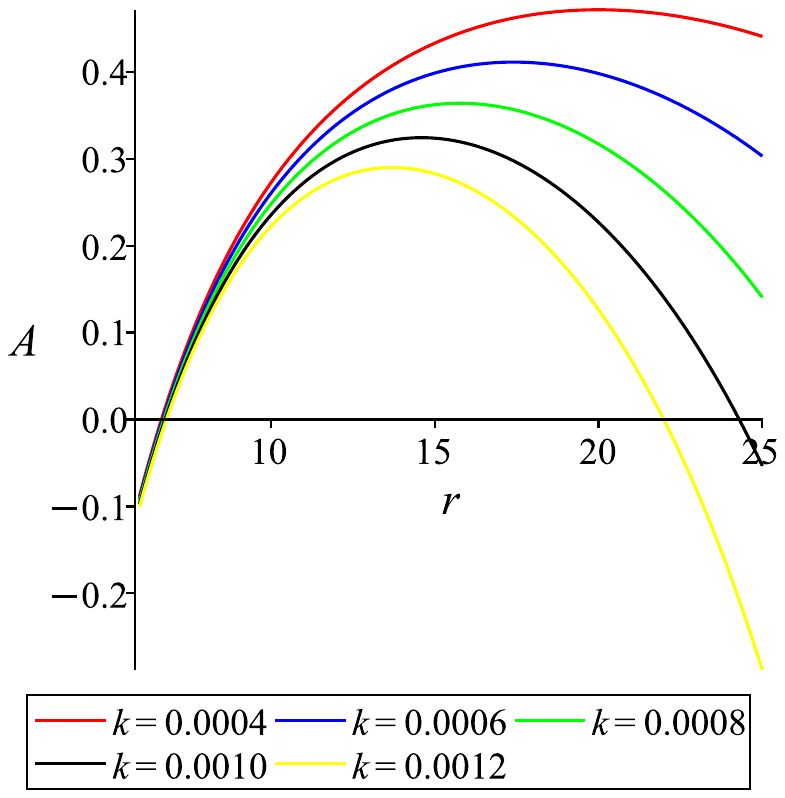} 
  \includegraphics[scale=0.4,trim = 1cm 9cm 2cm 1cm]{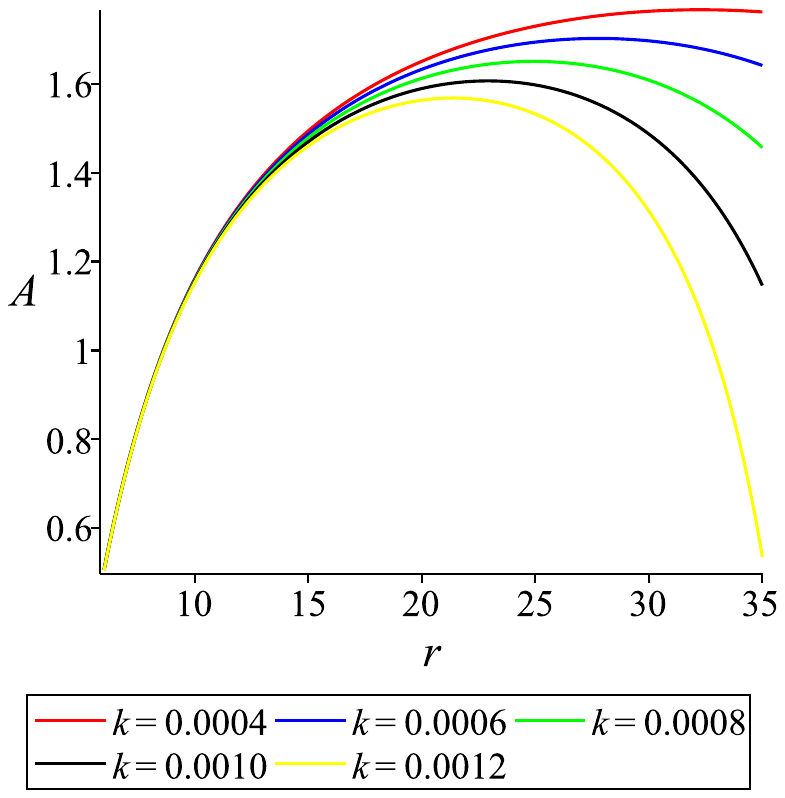}  
  \caption{The expression of $A$ for charged particle with the Hamiltonian defined in (\ref{HpotEl}) for $\varepsilon=1$ ({\it{left panel}}) and $\varepsilon=-1$ ({\it{right panel}}). The other numerical values used here are $M=1$ and $U=0.3$.} 
  \label{A12}
\end{figure}

Finally, one may notice that, depending on the particle's energy, the periapsis shift can be either positive (as in the left panel of the figure \ref{Electric}) or negative (as in the right panel of the figure \ref{Electric}).  The red and green dots correspond to the starting and ending points respectively.

\begin{figure}[H]
  \centering
 \includegraphics[scale=0.4,trim = 0cm 12cm 2cm 1cm]{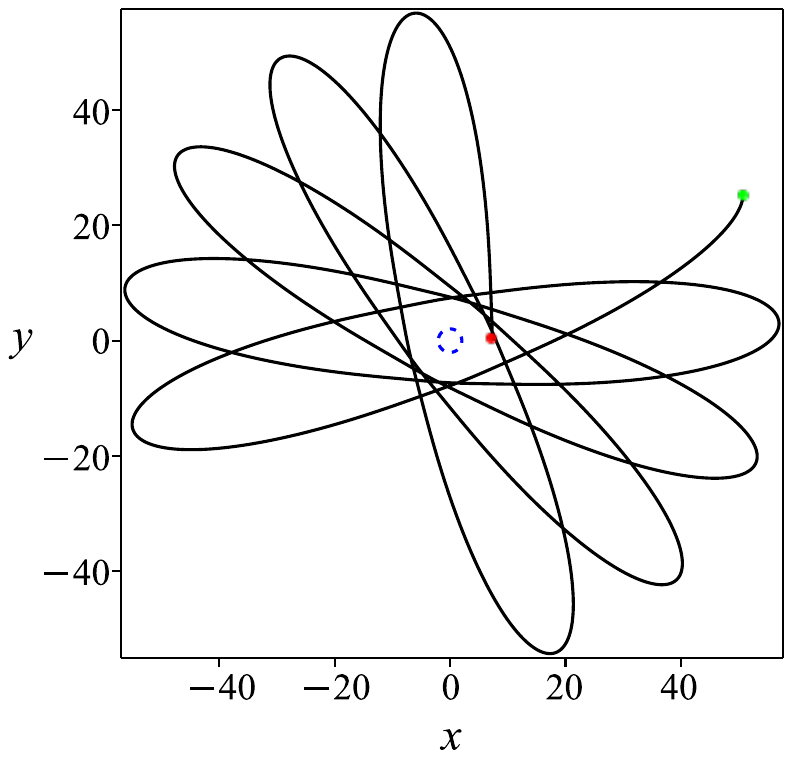}  
\includegraphics[scale=0.4,trim = 0cm 12cm 2cm 1cm]{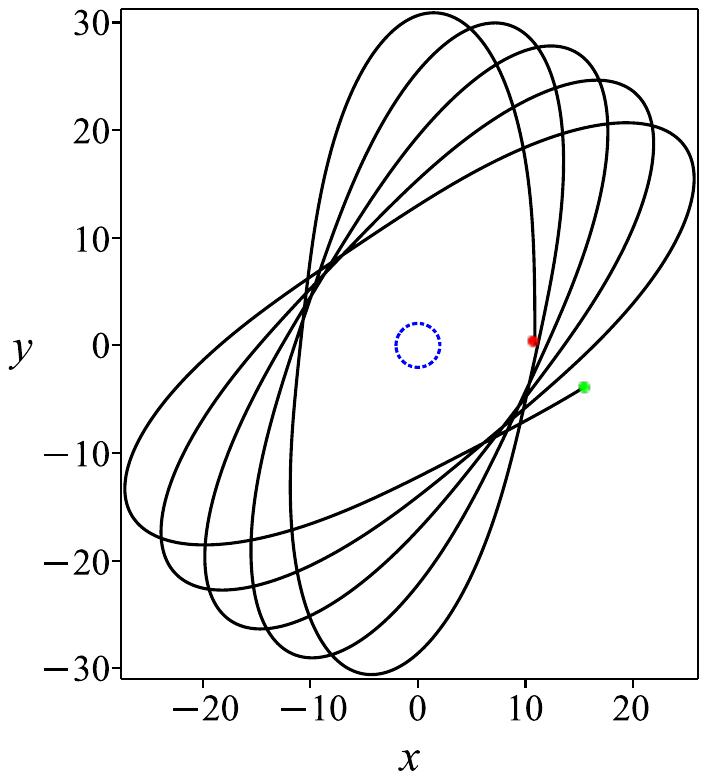}
  \caption{Parametric plot of a bounded trajectory of the charged particle trapped by the potential (\ref{potential}) for the metric function (\ref{metric23}). {\it The left panel.} The periastron shift is positive and the particle's energy is $E =0.26$. {\it The right panel}. The periastron shift is negative  and the particle's energy is $E =0.20$. The numerical values of the parameters are: $M=1$, $U=0.5$, $k=0.01$, $\varepsilon = -2$, $L = \sqrt{6}$.} 
  \label{Electric}
\end{figure}

\section{Conclusions}

 In the first part of the paper we introduced a new exact solution of Einstein’s field equations describing an electrically charged Kiselev black hole surrounded by a charged anisotropic “quintessence” fluid and we considered the motion of charged particles in this background.
For massive charged particles moving on timelike trajectories, we show that, depending on the model's parameters, the effective potential may allow bound orbits. This happens for large values of $U$ or/and $|w|$ and for the quintessence parameter below a critical value.
Similarly to the case of the Reissner-Nordstrom de Siter spacetime, the bound orbits show a noticeable deformation due to the electric interaction between the charged particle and the electrically charged black hole. A special attention is given to general conditions for the existence of circular trajectories.

The stability of the circular orbits can be analyzed by computing the Lyapunov exponent. Using a numerical approach, for $w=-2/3$ and values of $\varepsilon$ and $L$ for which charged particles move on stable circular orbits, it turns out that the Lyapunov exponent is very sensitive to the values of the parameters $U$ and $k$. As expected, once $k$ increases, the circular orbits become unstable. On the other hand, for particles with negative $\varepsilon$, when the charge of the black hole increases, the circular orbits are more stable. 

In the second part of this paper, namely in section 4 we considered the problem of the periapsis shift of the orbits of the charged particles in this background. As it is well-known, general relativistic effects in orbital motions around a central compact object imply the existence of a prograde periapsis shift of the orbit. However, there could be various other reasons that could lead to a retrograde periapsis shift: such a presence of dark matter around the compact object, or even the presence of naked singularities in the spacetimes describing the compact object. Our work provides one new example of charged spacetimes that might allow retrograde precessions for charged particles in these backgrounds. We found that for uncharged particles the periapsis shifts for bounded orbits is always prograde. However, for charged test particles the periapsis shifts can become retrograde. 

One important extension of the our results is a more in-depth study of the topological properties of circular orbits for charged or uncharged test particles in these backgrounds, along the lines of \cite{Song:2025vhw}, since the Kiselev black holes exhibit multiple horizons (black hole horizons and a cosmological horizon) similar to the de Sitter case. Another interesting extension of our work concerns the case of rotating black holes in scalar multipolar Universes, which are examples of rotating and charged black holes in presence of multipolar scalar fields \cite{Stelea:2025ppj}. Previous studies showed that the retrograde precession can occur for spaces with scalar fields, such as the \cite{Ota:2021mub}, \cite{Bambhaniya:2019pbr}.

Work on these matters is in progress and it will be presented elsewhere.


\newpage

\begin{thebibliography}
\footnotesize

\bibitem{LIGOScientific:2016aoc}
B.~P.~Abbott \textit{et al.} [LIGO Scientific and Virgo],
Phys. Rev. Lett. \textbf{116}, no.6, 061102 (2016)
doi:10.1103/PhysRevLett.116.061102
[arXiv:1602.03837 [gr-qc]].

\bibitem{LIGOScientific:2020ibl}
R.~Abbott \textit{et al.} [LIGO Scientific and Virgo],
Phys. Rev. X \textbf{11}, 021053 (2021)
doi:10.1103/PhysRevX.11.021053
[arXiv:2010.14527 [gr-qc]].

\bibitem{EventHorizonTelescope:2019dse}
K.~Akiyama \textit{et al.} [Event Horizon Telescope],
Astrophys. J. Lett. \textbf{875}, L1 (2019)
doi:10.3847/2041-8213/ab0ec7
[arXiv:1906.11238 [astro-ph.GA]].

\bibitem{SupernovaSearchTeam:1998fmf}
A.~G.~Riess \textit{et al.} [Supernova Search Team],
Astron. J. \textbf{116}, 1009-1038 (1998)
doi:10.1086/300499
[arXiv:astro-ph/9805201 [astro-ph]].

\bibitem{Tsujikawa:2013fta}
S.~Tsujikawa,
Class. Quant. Grav. \textbf{30}, 214003 (2013)
doi:10.1088/0264-9381/30/21/214003
[arXiv:1304.1961 [gr-qc]].

\bibitem{Brown:1997jv}
J.~D.~Brown and V.~Husain,
Int. J. Mod. Phys. D \textbf{6}, 563-573 (1997)
doi:10.1142/S0218271897000340
[arXiv:gr-qc/9707027 [gr-qc]].

\bibitem{Kiselev:2002dx}
V.~V.~Kiselev,
Class. Quant. Grav. \textbf{20}, 1187-1198 (2003)
doi:10.1088/0264-9381/20/6/310
[arXiv:gr-qc/0210040 [gr-qc]].

\bibitem{Azreg-Ainou:2017obt}
M.~Azreg-A{\"\i}nou, S.~Bahamonde and M.~Jamil,
Eur. Phys. J. C \textbf{77}, no.6, 414 (2017)
doi:10.1140/epjc/s10052-017-4969-4
[arXiv:1701.02239 [gr-qc]].

\bibitem{Konoplya:2019sns}
R.~A.~Konoplya,
Phys. Lett. B \textbf{795}, 1-6 (2019)
doi:10.1016/j.physletb.2019.05.043
[arXiv:1905.00064 [gr-qc]].

\bibitem{Zeng:2020vsj}
X.~X.~Zeng and H.~Q.~Zhang,
Eur. Phys. J. C \textbf{80}, no.11, 1058 (2020)
doi:10.1140/epjc/s10052-020-08656-7
[arXiv:2007.06333 [gr-qc]].

\bibitem{Bagchi:2025fxp}
B.~Bagchi and S.~Sen,
[arXiv:2510.07263 [gr-qc]].

\bibitem{Abdujabbarov:2015pqp}
A.~Abdujabbarov, B.~Toshmatov, Z.~Stuchl{\'\i}k and B.~Ahmedov,
Int. J. Mod. Phys. D \textbf{26}, no.06, 1750051 (2016)
doi:10.1142/S0218271817500511
[arXiv:1512.05206 [gr-qc]].

\bibitem{Dariescu:2022kof}
M.~A.~Dariescu, C.~Dariescu, V.~Lungu and C.~Stelea,
Phys. Rev. D \textbf{106}, no.6, 064017 (2022)
doi:10.1103/PhysRevD.106.064017
[arXiv:2206.12876 [gr-qc]].

\bibitem{Dariescu:2023twk}
M.~A.~Dariescu, V.~Lungu, C.~Dariescu and C.~Stelea,
Phys. Rev. D \textbf{109}, no.2, 024021 (2024)
doi:10.1103/PhysRevD.109.024021
[arXiv:2311.11356 [gr-qc]].

\bibitem{Jeong:2023hom}
S.~Jeong, B.~H.~Lee, H.~Lee and W.~Lee,
Phys. Rev. D \textbf{107}, no.10, 104037 (2023)
doi:10.1103/PhysRevD.107.104037
[arXiv:2301.12198 [gr-qc]].

\bibitem{Stelea:2018elx}
C.~Stelea, M.~A.~Dariescu and C.~Dariescu,
Phys. Rev. D \textbf{108}, no.8, 084034 (2023)
doi:10.1103/PhysRevD.108.084034
[arXiv:1810.02235 [gr-qc]].

\bibitem{Stelea:2018shx}
C.~Stelea, M.~A.~Dariescu and C.~Dariescu,
Phys. Rev. D \textbf{98}, no.12, 124022 (2018)
doi:10.1103/PhysRevD.98.124022
[arXiv:1810.03008 [gr-qc]].

\bibitem{Stelea:2018cgm}
C.~Stelea, M.~A.~Dariescu and C.~Dariescu,
Phys. Rev. D \textbf{97}, no.10, 104059 (2018)
doi:10.1103/PhysRevD.97.104059
[arXiv:1804.08075 [gr-qc]].

\bibitem{Weyl:1917gp}
H.~Weyl,
Annalen Phys. \textbf{54}, 117-145 (1917)
doi:10.1007/s10714-011-1310-7

\bibitem{GH} R. Gautreau and R. B. Hoffman, Nuovo Cimento B 16, 162 (1973).

\bibitem{Lemos:2009mr}
J.~P.~S.~Lemos and V.~T.~Zanchin,
Phys. Rev. D \textbf{80}, 024010 (2009)
doi:10.1103/PhysRevD.80.024010
[arXiv:0905.3553 [gr-qc]].

\bibitem{Wald}  R.~M.~Wald, General relativity, University of Chicago Press, Chicago, 1984.

\bibitem{Kormendy:2013dxa}
J.~Kormendy and L.~C.~Ho,
Ann. Rev. Astron. Astrophys. \textbf{51}, 511-653 (2013)
doi:10.1146/annurev-astro-082708-101811
[arXiv:1304.7762 [astro-ph.CO]].

\bibitem{Genzel2010}
 Genzel, R., Eisenhauer, F., \& Gillessen, S.\ 2010, Reviews of Modern Physics, 82, 4, 3121. doi:10.1103/RevModPhys.82.3121
 
\bibitem{EventHorizonTelescope:2022wkp}
K.~Akiyama \textit{et al.} [Event Horizon Telescope],
Astrophys. J. Lett. \textbf{930}, no.2, L12 (2022)
doi:10.3847/2041-8213/ac6674
[arXiv:2311.08680 [astro-ph.HE]].
 
\bibitem{EventHorizonTelescope:2022apq}
K.~Akiyama \textit{et al.} [Event Horizon Telescope],
Astrophys. J. Lett. \textbf{930}, no.2, L13 (2022)
doi:10.3847/2041-8213/ac6675
[arXiv:2311.08679 [astro-ph.HE]].

\bibitem{EventHorizonTelescope:2022wok}
K.~Akiyama \textit{et al.} [Event Horizon Telescope],
Astrophys. J. Lett. \textbf{930}, no.2, L14 (2022)
doi:10.3847/2041-8213/ac6429
[arXiv:2311.09479 [astro-ph.HE]].

\bibitem{EventHorizonTelescope:2022exc}
K.~Akiyama \textit{et al.} [Event Horizon Telescope],
Astrophys. J. Lett. \textbf{930}, no.2, L15 (2022)
doi:10.3847/2041-8213/ac6736
[arXiv:2311.08697 [astro-ph.HE]].

\bibitem{EventHorizonTelescope:2022urf}
K.~Akiyama \textit{et al.} [Event Horizon Telescope],
Astrophys. J. Lett. \textbf{930}, no.2, L16 (2022)
doi:10.3847/2041-8213/ac6672
[arXiv:2311.09478 [astro-ph.HE]].

\bibitem{EventHorizonTelescope:2022xqj}
K.~Akiyama \textit{et al.} [Event Horizon Telescope],
Astrophys. J. Lett. \textbf{930}, no.2, L17 (2022)
doi:10.3847/2041-8213/ac6756
[arXiv:2311.09484 [astro-ph.HE]].

\bibitem{EventHorizonTelescope:2024hpu}
K.~Akiyama \textit{et al.} [Event Horizon Telescope],
Astrophys. J. Lett. \textbf{964}, no.2, L25 (2024)
doi:10.3847/2041-8213/ad2df0

\bibitem{EventHorizonTelescope:2024rju}
K.~Akiyama \textit{et al.} [Event Horizon Telescope],
Astrophys. J. Lett. \textbf{964}, no.2, L26 (2024)
doi:10.3847/2041-8213/ad2df1


\bibitem{Vagnozzi:2022moj}
S.~Vagnozzi, R.~Roy, Y.~D.~Tsai, L.~Visinelli, M.~Afrin, A.~Allahyari, P.~Bambhaniya, D.~Dey, S.~G.~Ghosh and P.~S.~Joshi, \textit{et al.}
Class. Quant. Grav. \textbf{40}, no.16, 165007 (2023)
doi:10.1088/1361-6382/acd97b
[arXiv:2205.07787 [gr-qc]].

\bibitem{Ghez:2003rt}
A.~M.~Ghez, G.~Duchene, K.~Matthews, S.~D.~Hornstein, A.~Tanner, J.~Larkin, M.~Morris, E.~E.~Becklin, S.~Salim and T.~Kremenek, \textit{et al.}
Astrophys. J. Lett. \textbf{586}, L127-L131 (2003)
doi:10.1086/374804
[arXiv:astro-ph/0302299 [astro-ph]].

\bibitem{Eatough:2013nva}
R.~P.~Eatough, H.~Falcke, R.~Karuppusamy, K.~J.~Lee, D.~J.~Champion, E.~F.~Keane, G.~Desvignes, D.~H.~F.~M.~Schnitzeler, L.~G.~Spitler and M.~Kramer, \textit{et al.}
Nature \textbf{501}, 391-394 (2013)
doi:10.1038/nature12499
[arXiv:1308.3147 [astro-ph.GA]].

\bibitem{Gillessen:2008qv}
S.~Gillessen, F.~Eisenhauer, S.~Trippe, T.~Alexander, R.~Genzel, F.~Martins and T.~Ott,
Astrophys. J. \textbf{692}, 1075-1109 (2009)
doi:10.1088/0004-637X/692/2/1075
[arXiv:0810.4674 [astro-ph]].

\bibitem{GRAVITY:2020gka}
R.~Abuter \textit{et al.} [GRAVITY],
Astron. Astrophys. \textbf{636}, L5 (2020)
doi:10.1051/0004-6361/202037813
[arXiv:2004.07187 [astro-ph.GA]].


\bibitem{Fernando:2012ue}
S.~Fernando,
Gen. Rel. Grav. \textbf{44}, 1857-1879 (2012)
doi:10.1007/s10714-012-1368-x
[arXiv:1202.1502 [gr-qc]].

\bibitem{Ciou:2025ygb}
S.~Y.~Ciou, T.~Hsieh and D.~S.~Lee,
JCAP \textbf{05}, 086 (2025)
doi:10.1088/1475-7516/2025/05/086
[arXiv:2503.12911 [gr-qc]].

\bibitem{Olivares:2011xb}
M.~Olivares, J.~Saavedra, J.~R.~Villanueva and C.~Leiva,
Mod. Phys. Lett. A \textbf{26}, 2923-2950 (2011)
doi:10.1142/S0217732311037261
[arXiv:1101.0748 [gr-qc]].

\bibitem{Pugliese:2010ps}
D.~Pugliese, H.~Quevedo and R.~Ruffini,
Phys. Rev. D \textbf{83}, 024021 (2011)
doi:10.1103/PhysRevD.83.024021
[arXiv:1012.5411 [astro-ph.HE]].

\bibitem{Battista:2022krl}
E.~Battista and G.~Esposito,
Eur. Phys. J. C \textbf{82}, no.12, 1088 (2022)
doi:10.1140/epjc/s10052-022-11070-w
[arXiv:2202.03763 [gr-qc]].

\bibitem{Garnier:2025jnp}
A.~Garnier and E.~Battista,
Eur. Phys. J. C \textbf{85}, no.3, 284 (2025)
doi:10.1140/epjc/s10052-025-13957-w
[arXiv:2502.10053 [gr-qc]].

\bibitem{Harada:2022uae}
T.~Harada, T.~Igata, H.~Saida and Y.~Takamori,
Int. J. Mod. Phys. D \textbf{32}, no.15, 2350098 (2023)
doi:10.1142/S0218271823500980
[arXiv:2210.07516 [gr-qc]].

\bibitem{Igata:2022rcm}
T.~Igata, T.~Harada, H.~Saida and Y.~Takamori,
Int. J. Mod. Phys. D \textbf{32}, no.16, 2350105 (2023)
doi:10.1142/S0218271823501055
[arXiv:2202.00202 [gr-qc]].

\bibitem{Katsumata:2025jrf}
A.~Katsumata and T.~Harada,
[arXiv:2507.04280 [gr-qc]].

\bibitem{Igata:2022nkt}
T.~Igata and Y.~Takamori,
Phys. Rev. D \textbf{105}, no.12, 124029 (2022)
doi:10.1103/PhysRevD.105.124029
[arXiv:2202.03114 [gr-qc]].

\bibitem{Bini:2005dy}
D.~Bini, F.~De Paolis, A.~Geralico, G.~Ingrosso and A.~Nucita,
Gen. Rel. Grav. \textbf{37}, 1263-1276 (2005)
doi:10.1007/s10714-005-0109-9
[arXiv:gr-qc/0502062 [gr-qc]].

\bibitem{Bambhaniya:2021ybs}
P.~Bambhaniya, D.~Dey, A.~B.~Joshi, P.~S.~Joshi, D.~N.~Solanki and A.~Mehta,
Phys. Rev. D \textbf{103}, no.8, 084005 (2021)
doi:10.1103/PhysRevD.103.084005
[arXiv:2101.03865 [gr-qc]].

\bibitem{Bambhaniya:2019pbr}
P.~Bambhaniya, A.~B.~Joshi, D.~Dey and P.~S.~Joshi,
Phys. Rev. D \textbf{100}, no.12, 124020 (2019)
doi:10.1103/PhysRevD.100.124020
[arXiv:1908.07171 [gr-qc]].

\bibitem{Bambhaniya:2025xmu}
P.~Bambhaniya, M.~J.~Vyas, P.~S.~Joshi and E.~M.~de Gouveia Dal Pino,
Phys. Dark Univ. \textbf{48}, 101949 (2025)
doi:10.1016/j.dark.2025.101949
[arXiv:2501.11232 [gr-qc]].

\bibitem{Dey:2019fpv}
D.~Dey, P.~S.~Joshi, A.~Joshi and P.~Bambhaniya,
Int. J. Mod. Phys. D \textbf{28}, no.14, 1930024 (2019)
doi:10.1142/S0218271819300246
[arXiv:2101.06001 [gr-qc]].

\bibitem{Ota:2021mub}
K.~Ota, S.~Kobayashi and K.~Nakashi,
Phys. Rev. D \textbf{105}, no.2, 024037 (2022)
doi:10.1103/PhysRevD.105.024037
[arXiv:2110.07503 [gr-qc]].

\bibitem{Arguelles:2021jtk}
C.~R.~Arg{\"u}elles, M.~F.~Mestre, E.~A.~Becerra-Vergara, V.~Crespi, A.~Krut, J.~A.~Rueda and R.~Ruffini,
Mon. Not. Roy. Astron. Soc. \textbf{511}, no.1, L35-L39 (2022)
doi:10.1093/mnrasl/slab126
[arXiv:2109.10729 [astro-ph.GA]].

\bibitem{Atamurotov:2022nim}
F.~Atamurotov, I.~Hussain, G.~Mustafa and K.~Jusufi,
Eur. Phys. J. C \textbf{82}, no.9, 831 (2022)
doi:10.1140/epjc/s10052-022-10782-3
[arXiv:2209.01652 [gr-qc]].

\bibitem{Gohain:2024piy}
M.~M.~Gohain, K.~Bhuyan, R.~Borgohain, T.~Gogoi, K.~Bhuyan and P.~Phukon,
Nucl. Phys. B \textbf{1018}, 117073 (2025)
doi:10.1016/j.nuclphysb.2025.117073
[arXiv:2412.06252 [gr-qc]].

\bibitem{Gao:2022ybw}
C.~Gao, D.~Chen, C.~Yu and P.~Wang,
Phys. Lett. B \textbf{833}, 137343 (2022)
doi:10.1016/j.physletb.2022.137343
[arXiv:2204.07983 [gr-qc]].


\bibitem{Stuchlik:2021guq}
Z.~Stuchl{\'\i}k and J.~Vrba,
Eur. Phys. J. Plus \textbf{136}, no.11, 1127 (2021)
doi:10.1140/epjp/s13360-021-02078-4
[arXiv:2110.10569 [gr-qc]].

\bibitem{Stelea:2023yqo}
C.~Stelea, M.~A.~Dariescu and C.~Dariescu,
Phys. Lett. B \textbf{847}, 138275 (2023)
doi:10.1016/j.physletb.2023.138275
[arXiv:2309.13651 [gr-qc]].

\bibitem{Avalos-Vargas:2012jja}
A.~Avalos-Vargas and G.~Ares De Parga,
Eur. Phys. J. Plus \textbf{127}, 155 (2012)
doi:10.1140/epjp/i2012-12155-2

\bibitem{Stuchlik:2002tj}
Z.~Stuchlik and S.~Hledik,
Acta Phys. Slov. \textbf{52}, no.5, 363-407 (2002)
[arXiv:0803.2685 [gr-qc]].

\bibitem{Sau:2020xau}
S.~Sau, I.~Banerjee and S.~SenGupta,
Phys. Rev. D \textbf{102}, no.6, 064027 (2020)
doi:10.1103/PhysRevD.102.064027
[arXiv:2004.02840 [gr-qc]].
 
\bibitem{Song:2025vhw}
Y.~Song, J.~Li, Y.~Cen, K.~Diao, X.~Zhao and S.~Shi,
[arXiv:2504.05061 [gr-qc]].

\bibitem{Stelea:2025ppj}
C.~Stelea, M.~A.~Dariescu and V.~Lungu,
[arXiv:2501.11807 [gr-qc]].

\end{thebibliography}
\end{document}